\documentclass[12pt]{iopart}
\usepackage{amsfonts}
\usepackage{bm}
\usepackage{amssymb}
\usepackage{epsfig}

\begin{document}
\title{Control of stochasticity in magnetic field lines}

\author{C Chandre$^1$, M Vittot$^1$, G Ciraolo$^2$, Ph Ghendrih$^2$, R Lima$^1$}
\address{$^1$ Centre de Physique Th\'eorique\footnote{Unit\'e Mixte de Recherche (UMR 6207) du CNRS, et des universit\'es Aix-Marseille I, Aix-Marseille II et du Sud Toulon-Var. Laboratoire affili\'e \`a la
FRUMAM (FR 2291).}, CNRS Luminy, Case 907, F-13288 Marseille Cedex 9,
France}
\address{$^2$ Association Euratom-CEA, DRFC/DSM/CEA, CEA Cadarache,
F-13108 St. Paul-lez-Durance Cedex, France}
\eads{\mailto{chandre@cpt.univ-mrs.fr}, \mailto{vittot@cpt.univ-mrs.fr}, \mailto{guido.ciraolo@cea.fr}, \mailto{philippe.ghendrih@cea.fr}, \mailto{lima@cpt.univ-mrs.fr} }

\begin{abstract}
We present a method of control which is able to create barriers to magnetic field line diffusion by a small modification of the magnetic perturbation. This method of control is based on a {\em localized} control of chaos in Hamiltonian systems. The aim is to modify the perturbation (of order $\varepsilon$) {\em locally} by a small control term (of order $\varepsilon^2$) which creates invariant tori acting as barriers to diffusion for Hamiltonian systems with two degrees of freedom. The location of the invariant torus is enforced in the vicinity of the chosen target (at a distance of order $\varepsilon$ due to the angle dependence). Given the importance of confinement in magnetic fusion devices, the method is applied to two examples with a loss of magnetic confinement. In the case of locked tearing modes, an invariant torus can be restored that aims at showing the current quench and therefore the generation of runaway electrons. In the second case, the method is applied to the control of stochastic boundaries allowing one to define a transport barrier within the stochastic boundary and therefore to monitor the volume of closed field lines.      
\end{abstract}


\maketitle

\section{Introduction}
\label{sec:1}
Controlling chaotic transport is a key challenge in many branches
of physics like for instance, in particle accelerators, free
electron lasers or in magnetically confined fusion
plasmas~\cite{review1,review2,cary82,hans84,cary86,hans94,huds97,
huds98,spat98,reim01,huds01,huds02,huds04}. It covers a large variety of strategies. On the one
hand, one can aim at recovering near integrable systems while on
the other hand, one can request the existence of a boundary with
strongly reduced transport that will act as some transport
barrier. Another feature that will prove to be essential when
implementing the control scheme is the relative cost of such a
control device with respect to its merits. The cost that is
considered here will be characterized by the magnitude and
complexity of the Fourier spectrum of the control function with
respect to that generating the spurious stochastic transport. In
the case of systems such that it is essential to control a global
transport property without significantly altering the original
system under investigation nor its overall chaotic structure, one
can restrict the control scheme to that of restoring a local
transport barrier. In accordance with the latter idea, the control
strategy that we develop here is based on building barriers by
adding a small apt perturbation which is localized in phase space,
hence confining all the trajectories. This local control strongly
reduces the cost of the control scheme since only a small fraction
of the phase space is modified. The counterpart is that the impact
of such a localized reduction in transport must be efficient
enough to be meaningful regarding the control strategy.

In the case of particle accelerators, free electron lasers, and magnetic fusion, the magnetic field is designed to confine the charged particles. Hamiltonian dynamics allows one to
describe the motion of the charged particles in such a magnetic
field, and the confinement volumes readily take the form of
invariant tori. The Hamiltonian formalism appears to be
particularly well suited to investigate the control of such
systems. In this article, we consider the class of Hamiltonian
systems that can be written in the form $H=H_0+\varepsilon V$
i.e.\ an integrable Hamiltonian $H_0$ (with action-angle
variables) plus a perturbation $\varepsilon V$. Provided the
perturbation is not too small, the explicit dependence of
$\varepsilon V$ on the angle variables breaks the invariance of
the action variables and in a generic situation yields chaotic
transport. The idea for controlling this transport is to modify
the perturbation {\em slightly} and {\em locally} and create
regular structures (like invariant tori).

In the theory of magnetic confinement, the particle trajectories are characterized by three invariants, namely the magnetic moment, that is in fact an adiabatic invariant, the kinetic moment along the vertical axis which is the axis of symmetry of the axisymmetric equilibrium, and the particle energy (for a collisionless plasma). In steady state solutions, the various quantities of relevance for transport must solely depend on these invariants and not on the associated phases. When considering magnetic field lines, this six dimensional problem of the particle motion in phase space, can be reduced to a two dimensional one. The time variable then appears to be a toroidal angle, while the invariant is the toroidal flux and the Hamiltonian, the poloidal flux.  The conjugate angle of the toroidal flux is found to be a poloidal angle. In the case of a magnetic equilibrium, the poloidal flux is solely a function of the action variable, the toroidal flux, both being labels of the magnetic surfaces, i.e.\ invariant tori. In this approach, and for an axisymmetric equilibrium, the energy, namely the poloidal flux, cannot depend on the angle since the latter depends linearly on the toroidal angle. It is therefore important that the unperturbed Hamiltonian for an axisymmetric equilibrium does not depend on the angle.

In Refs.~\cite{guido1,guido2,michel}, an explicit method of
control was provided for generating $f$ such that the controlled
Hamiltonian $H_0+\varepsilon V+\varepsilon^2 f$ is integrable. We
point out that the control term we constructed is of order
$\varepsilon^2$, i.e.\ much smaller than the perturbation which is
of order $\varepsilon$. This method of control has been applied experimentally to increase the kinetic coherence of an electron beam in a Traveling Wave Tube~\cite{prl05}. One possible drawback of this approach is that
the control term depends on all the variables and has to be applied on all the phase space. Here we
provide a method to construct control terms $f$ with a finite
support in phase space, i.e.\ localized in phase space, such that
the controlled Hamiltonian $H_c=H_0+\epsilon V+ \varepsilon^2 f$
has invariant tori whose explicit expressions are known.
For Hamiltonian systems with two degrees of freedom, these
invariant tori act as barriers to diffusion in phase space. 

In the present paper, this original method is applied to the
control of chaotic transport in fusion plasmas. However, these
examples are sufficiently general so that one can readily
transpose the approach to other fields of physics. Let us consider
two examples where the control of chaotic transport appears to be
of particular interest.

In a first example, one addresses the issue of large scale
stochastic transport between two resonant surfaces. In tokamaks,
this situation is met when two tearing modes develop~\cite{samain} and when
their amplitude is large enough to satisfy the so-called overlap
criterion proposed by Chirikov~\cite{chir79}. This leads to chaotic transport
from one resonant surface to the other. The control scheme can
target two results, first to maintain a closed magnetic surface (an invariant torus)
so that confinement can be sustained
until some other action is taken to suppress the tearing modes,
second, should the tearing mode interaction have led to a
pre-disruptive phase, to generate a transport barrier sufficiently
robust to slow the current quench and provide conditions for a
controlled disruption with no generation of runaway electrons~\cite{iter}.

In a second example, one considers the situation of an external
modification of the magnetic equilibrium, for instance in the case
of an ergodic divertor~\cite{finken} where specific coils are implemented to
generate chaotic transport in the outermost magnetic surfaces. A
similar situation can be met in stellarators where the magnetic
equilibrium can be such that a series of resonances in the plasma
boundary creates a region of chaotic transport. In these
cases, it can be of interest to control this chaotic boundary
layer by restoring a closed magnetic surface at a given radius. In
the latter case, one should consider a control scheme with similar
properties to that generating the boundary layer with stochastic
transport. This would ensure that a similar coil set to that
generating the perturbation of the magnetic equilibrium can be
used for the control purpose, hence ensuring some effectiveness in
terms of cost and feasibility.

In Sec.~\ref{sec:2}, we provide an explicit construction of the localized
control term. The equation of the created invariant
torus is given explicitly. In Sec.~\ref{sec:3}, the formula of the control term
for magnetic field lines is derived as well as the explicit formula for the magnetic
surface which has been created by the localized control term. In Sec.~\ref{sec:4},
the two examples of magnetic perturbations are considered. It is shown on these examples
that a small and apt modification of the poloidal flux is able to create a robust
magnetic surface which prevents the diffusion of magnetic field lines. In the Appendix, we give a proof of existence of the control term.

\section{Control of chaos in Hamiltonian systems}
\label{sec:2}

In this section we present a control method of Hamiltonian
systems which is directly applicable to magnetic field lines. This
version of localized control follows the one
developed in Refs.~\cite{guido4,vitt04}. In order to understand the
procedure to build the control scheme, let us first analyze the
properties of a specific class of Hamiltonian systems that will
provide the background for the control strategy. In the following
${\bf A}$ refers to the action variables and ${\bm\theta}$ to the
angle variables in a phase space of dimension $2L$ (hence both vectors are
of dimension $L$). Let us consider Hamiltonian systems of the form
$H({\bf A},{\bm\theta})=H_0({\bf A})+\varepsilon V({\bf A},{\bm\theta})$. Using a suitable expansion, these Hamiltonian systems take the form
\begin{equation}
\label{eqn:ham}
    H({\bf A},{\bm\theta})={\bm \omega}\cdot {\bf A}+ \varepsilon v({\bm \theta}) +w({\bf A},{\bm\theta}),
\end{equation}
where $({\bf A},{\bm\theta})\in {\mathbb R}^L\times {\mathbb
T}^L$, ${\mathbb T}$ being an angle space (torus) $[-\pi, \pi[$ in
a standard approach. The three contributions to $H({\bf
A},{\bm\theta})$ are respectively $ H_{0}({\bf A})={\bm
\omega}\cdot {\bf A}$, the main term governing the integrable
motion in the vicinity of ${\bf A}={\bf 0}$, a perturbation $\varepsilon
v({\bm \theta})$ and a higher order term in $\bf A$, $w({\bf
A},{\bm\theta})$ which can be written in the form
$$
w({\bf A},{\bm\theta})=\varepsilon {\bf w}_1({\bm\theta})\cdot {\bf A}+ w_2({\bf A},{\bm\theta}),
$$
where $w_2$ is quadratic in the actions, i.e.\ $w_2({\bf 0},{\bm\theta})=0$ and $\partial_{\bf A}w_2({\bf 0},{\bm\theta})={\bf 0}$.

The vector $\bm\omega$ contains the frequencies of the
quasi-periodic motion defined by $\dot{\bf A}={\bf 0}$ and
$\dot{\bm\theta}=\bm\omega$. Furthermore, $\bm\omega$ is a
non-resonant vector of ${\mathbb R}^L$, i.e.\ there is no non-zero
${\bf k}\in {\mathbb Z}^L$ such that ${\bm\omega}\cdot {\bf k}=0$.
Without restricting the computation of the control term, one can
assume that $w({\bf 0},{\bm\theta})=0$ for all
${\bm\theta}\in{\mathbb T}^L$ and that $\int_{{\mathbb T}^L}
v({\bm\theta})d^L{\bm\theta}=0$ (the average of $v$ is set to zero). We consider a region near ${\bf
A}={\bf 0}$. We notice that for $\varepsilon=0$ and for any $w$
not necessarily small, the Hamiltonian $H$ has an invariant torus
at ${\bf A}={\bf 0}$. The controlled Hamiltonian we construct is
given by
\begin{equation}
\label{eqn:gene}
H_c({\bf A},{\bm\theta})={\bm \omega}\cdot {\bf A}+\varepsilon v({\bm \theta}) +w({\bf A},{\bm\theta}) +  \varepsilon^2 f({\bm \theta})\Omega({\bf A},{\bm \theta}),
\end{equation}
where $\Omega$ is a smooth characteristic function of a region around a targeted invariant torus.  We wish to stress that the actual control term $\varepsilon^2 f$ that we construct only
 depends on the angle variables ${\bm\theta}$ and is of order $\varepsilon^2$ (see Appendix).
We prove that the control term is ~:
\begin{equation}
\label{eqn:CT}
    f({\bm\theta})=-\varepsilon^{-2} w(-\varepsilon \Gamma \partial_{\bm\theta} v,{\bm\theta}),
\end{equation}
where $\partial_{\bm\theta}v$ denotes the first derivatives of $v$
with respect to ${\bm\theta}$~: $ \partial_{\bm\theta}v=\sum_{{\bf
k}\in{\mathbb Z}^L}i{\bf k} v_{\bf k} {\mathrm e}^{i{\bf k}\cdot
{\bm\theta}}$ and where the linear operator $\Gamma$ is the
inverse operator of the convective shift
$\dot{{\bm\theta}}\cdot \partial_{\bm\theta}$ relative to the
unperturbed dynamics governed by $H_0({\bf A})$. Its explicit form
acting on a function $v({\bm\theta})=\sum_{{\bf k}\in{\mathbb
Z}^L}v_{\bf k} {\mathrm e}^{i{\bf k}\cdot {\bm\theta}}$ is~:
$$
\Gamma v({\bm\theta})=\sum_{{\bf k}\in{\mathbb Z}^L\setminus \{ {\bf 0}\} }
\frac{v_{\bf k}}
{i{\bm \omega}\cdot {\bf k}} {\mathrm e}^{i{\bf k}\cdot {\bm\theta}},
$$
 The operators $\Gamma$ and $\partial_{\bm\theta}$
commute so that $\partial_{\bm\theta}(\Gamma v)=\Gamma(\partial_{\bm\theta} v)$. 

There is a significantly large freedom in choosing the function $\Omega$. It is sufficient to have $\Omega( {\bf A},{\bm\theta})=1$ for $\Vert {\bf A}\Vert \leq \varepsilon$. For instance, $\Omega( {\bf A},{\bm\theta})=1$ would be a possible and simpler choice, however representing a long-range control since the control term $f({\bm\theta})$ would be applied on all phase space. On the opposite way, we can design a function $\Omega$ such that the control is localized around the created invariant torus~:
We denote ${\mathcal T}_\alpha$ and ${\mathcal T}_\beta$ two neighborhoods (in phase
space) of the targeted invariant torus such that ${\mathcal T}_\alpha \subset {\mathcal T}_\beta$. The characteristic
function is chosen such that $\Omega(x)=1$ if $x\in {\mathcal T}_\alpha$ and $\Omega(x)=0$ if $x\notin {\mathcal T}_\beta$,
 and $\Omega$ is smooth on all phase space. We choose the
characteristic function $\Omega$ to be
$$
    \Omega({\bf A},{\bm \theta})=\Omega_{loc}(\Vert {\bf A}+\varepsilon \Gamma \partial_{\bm\theta} v\Vert),
$$
where $\Omega_{loc}(x)=1$ if $x\leq \alpha$ and $\Omega_{loc}(x)=0$ if $x\geq \beta$ and, e.g., a
polynomial or another function for $x\in ]\alpha,\beta[$ such that $\Omega_{loc}$ is smooth on ${\mathbb R}^+$. The main advantage of this step function $\Omega$ is that the control needs less energy (only in the part of phase space where the control is localized) and also it does not change the other part of phase space.

In these cases, we prove that $H_c$ given by Eq.~(\ref{eqn:gene}) has an invariant torus located at ${\bf A}=-\varepsilon \Gamma \partial_{\bm\theta} v$.
For Hamiltonian systems with two degrees of freedom, such an invariant torus acts as a barrier to diffusion. For the construction of the control term, we notice that we do not require that the quadratic part of $w$ is small in order to have a control term of order $\varepsilon^2$. Moreover, if the lowest order in powers of ${\bf A}$ of $w$ is $n\geq 2$ then the control term is of order $\varepsilon^{n}$.

In order to derive the expression of the control term, we consider the canonical transformation generated by
$$
F({\bf A}',{\bm\theta})={\bm\theta}\cdot {\bf A}'-\varepsilon\Gamma v({\bm\theta}),
$$
which is a translation in action by $\varepsilon\Gamma\partial_{\bm\theta}v$, i.e.\
\ $({\bf A}',{\bm\theta}')=({\bf A}+\varepsilon\Gamma\partial_{\bm\theta}v,{\bm\theta})$.
The Hamiltonian $H_c$ is mapped onto
$$
\tilde{H_c}({\bf A}',{\bm\theta}')={\bm \omega}\cdot {\bf A}' +w({\bf A}'-
\varepsilon\Gamma \partial_{\bm\theta} v,{\bm\theta}') +  \varepsilon^2 f({\bm \theta}')\Omega(\Vert {\bf A}'\Vert ).
$$
The translation in action is such that the contribution
$\varepsilon v$ is canceled out since ${\bm\omega} \cdot \Gamma
\partial_{\bm\theta}v=v$. The action ${\bf A}'={\bf 0}$ is found to be conserved by the
flow of $\tilde{H}_c$ since
$$
\frac{d{\bf A}'}{dt}=-\partial_{\bm\theta} w ({\bf
A}'-\varepsilon\Gamma \partial_{\bm\theta}
v,{\bm\theta}')+\varepsilon(\Gamma \partial^2_{\bm\theta} v)
\partial_{\bf A} w ({\bf A}'- \varepsilon\Gamma
\partial_{\bm\theta} v,{\bm\theta}')- \varepsilon^2
(\partial_{\bm\theta}f) \Omega(\Vert {\bf A}'\Vert ),
$$
and since the control term $f({\bm\theta})$ defined in Eq.~(\ref{eqn:CT}),
is such that ~:
$$
\partial_{\bm\theta}f=-\varepsilon^{-2}\partial_{\bm\theta} w (-\varepsilon\Gamma \partial_{\bm\theta} v,{\bm\theta})+\varepsilon^{-1}(\Gamma \partial^2_{\bm\theta} v) \partial_{\bf A} w (-\varepsilon\Gamma \partial_{\bm\theta} v,{\bm\theta}).
$$
As a consequence, in the domain such that
$\Omega(\Vert {\bf A}'\Vert )= 1$, we find ${\bf A}'={\bf 0}$ implies $d{\bf A}'/dt={\bf 0}$, and
consequently ${\bf A}'={\bf 0}$ is an invariant surface for $\tilde{H_c}$.

\section{Magnetic field lines}
\label{sec:3}

The magnetic field line dynamics in a toroidal geometry can be written in a Hamiltonian form~\cite{booz83,cary83,Bwhit89}
\begin{eqnarray*}
    && \frac{d\theta}{d\varphi}=\frac{\partial H}{\partial \psi},\\
    && \frac{d\psi}{d\varphi}=-\frac{\partial H}{\partial\theta},
\end{eqnarray*}
where $\varphi$ which plays the role of effective time is the
toroidal angle, $\psi$ is the normalized toroidal flux and $H$ is
the poloidal flux. The poloidal angle $\theta$ is the conjugate
variable to the action $\psi$. We consider the following class of
Hamiltonian systems~:
\begin{equation}
\label{eqn:Hms}
    H(\psi,\theta,\varphi)=H_0(\psi)+\varepsilon H_1(\psi,\theta,\varphi).
\end{equation}
We denote $Q(\psi)=H'_0(\psi)$. The quantity $q=1/Q$ is the safety
factor. For $\varepsilon = 0$,
we recover the unperturbed magnetic equilibrium such that ${d\psi}
/ {d\varphi}=0$, $\psi = \psi_0 $ are invariant tori also
characterized by the rotational transform
$q(\psi_0)={d\varphi}/{d\theta}$. We select a magnetic surface by
its unperturbed action $\psi=\psi_0$ where one wants to build a
barrier to diffusion. We expand $H_0$ into~:
$$
H_0(\psi)=Q(\psi_0)(\psi-\psi_0) +\sum_{l=1}^\infty \frac{1}{(l+1)!}Q^{(l)}(\psi_0) (\psi-\psi_0)^{l+1}.
$$
We denote $\omega=Q(\psi_0)=1/q(\psi_0)$ the winding ratio of the selected magnetic surface, and $Q^{(l)}$ denotes the $l$-th derivative of $Q$. Following the notation of Sec.~\ref{sec:2}, we have
$$
v(\theta,\varphi)=H_1(\psi_0,\theta,\varphi),
$$
and
$$
w(\psi,\theta,\varphi)=H_0(\psi)-Q(\psi_0)(\psi-\psi_0)+\varepsilon [H_1(\psi,\theta,\varphi)-H_1(\psi_0,\theta,\varphi) ].
$$
We expand $w$~:
$$
w(\psi,\theta,\varphi)=\varepsilon \partial_\psi H_1(\psi_0,\theta,\varphi)(\psi-\psi_0)+
\sum_{l=1}^\infty \frac{1}{(l+1)!}\left[ Q^{(l)}(\psi_0)+\varepsilon \partial_\psi^{l+1} H_1 (\psi_0,\theta,\varphi)\right](\psi-\psi_0)^{l+1}.
$$
We notice that $w(\psi_0,\theta,\varphi)=0$ and that $\partial_\psi w(\psi_0,\theta,\varphi)=\varepsilon \partial_\psi H_1(\psi_0,\theta,\varphi)$ is of order $\varepsilon$. Thus this Hamiltonian satisfies the requirements to construct a localized control term of order $\varepsilon^2$.\\
Following Refs.~\cite{abdu98,abdu99,abdu04}, the Fourier expansion of $H_1$  is given by
$$
H_1(\psi_0,\theta,\varphi)=\sum_{m,n} H_{mn}(\psi_0)\cos (m\theta-n\varphi+\chi_{mn}),
$$
for some constant phases $\chi_{mn}$.
In order to apply the control procedure described in the previous section, we consider that $\varphi$ is an angle variable and $E$ its conjugate action. In this way, we map the ``non-autonomous'' Hamiltonian system with 1.5 degrees of freedom into an ``autonomous'' Hamiltonian of the form~(\ref{eqn:ham}) with two degrees of freedom where the actions are ${\bf A}=(\psi-\psi_0,E)$ and the angles are ${\bm\theta}=(\theta,\varphi)$. Hamiltonian~(\ref{eqn:Hms}) has the form~(\ref{eqn:ham}) with ${\bm\omega}=(\omega,1)$.

The control term is given by
\begin{eqnarray}
     &&f(\theta,\varphi)=\left[ \partial_\psi H_1(\psi_0,\theta,\varphi)\right] \Gamma \partial_{\theta} H_1(\psi_0,\theta,\varphi)\nonumber \\
    && \quad -\sum_{l=1}^{\infty}\frac{(-\varepsilon)^{l-1}}{(l+1)!}\left( Q^{(l)}(\psi_0)+\varepsilon \partial_\psi^{l+1} H_1(\psi_0,\theta,\varphi)\right)
     \left(  \Gamma \partial_{\theta} H_1(\psi_0,\theta,\varphi)  \right) ^{l+1}, \label{eqn:ctl}
\end{eqnarray}
where
\begin{equation}
\label{eqn:GaD}
\Gamma \partial_{\theta} H_1(\psi_0,\theta,\varphi)=\sum_{m,n} \frac{m H_{mn}(\psi_0)}{m\omega -n}\cos (m\theta-n\varphi+\chi_{mn}).
\end{equation}
Using this control term, the controlled Hamiltonian has the invariant torus whose equation is
\begin{equation}
\label{eqn:surf}
\psi=\psi_0-\varepsilon \Gamma\partial_\theta H_1(\psi_0,\theta,\varphi),
\end{equation}
The difference between $\psi$ and $\psi_0$, of order
$\varepsilon$, is a function of $\psi_0$ both via the dependence
of $H_1$ on $\psi_0$ and that of the operator $\Gamma$ with
respect to the frequency $\omega=Q(\psi_0)$. The dominant term of
the control is given by considering only $l=1$ in
Eq.~(\ref{eqn:ctl}), which leads to
\begin{equation}
\label{eqn:f2}
f_2(\theta,\varphi)=\left[ \partial_\psi H_1(\psi_0,\theta,\varphi)\right] \Gamma\partial_\theta H_1(\psi_0,\theta,\varphi)- \frac{Q'(\psi_0)}{2}\left(  \Gamma\partial_\theta H_1(\psi_0,\theta,\varphi)  \right)^2,
\end{equation}
where $\Gamma\partial_{\theta} H_1$ is given by Eq.~(\ref{eqn:GaD}). The full control term is given as a series in $\varepsilon$~:
$$
\varepsilon^2 f=\sum_{s=2}^\infty \varepsilon^s f_s,
$$
where $f_s$ is given by
\begin{equation}
\label{eqn:fs}
f_s(\theta,\varphi)=\frac{(-1)^s}{(s-1)!}(\partial_\psi^{s-1} H_1) (\Gamma \partial_\theta H_1)^{s-1}
+\frac{(-1)^{s+1}}{s!} Q^{(s-1)}(\psi_0)(\Gamma \partial_\theta H_1)^s,
\end{equation}
for $s\geq 2$.

In summary, the controlled Hamiltonian we consider is
\begin{equation}
\label{eqn:HmsC}
    H_c=\int \frac{d\psi}{q(\psi)}+\varepsilon H_1(\psi,\theta,\varphi)+ \varepsilon^2 f(\theta,\varphi) \Omega(\Vert \psi-\psi_0+\varepsilon \Gamma \partial_\theta H_1(\psi_0,\theta,\varphi)\Vert),
\end{equation}
where $f$ is given by Eq.~(\ref{eqn:ctl}) for the complete control
term, or by Eq.~(\ref{eqn:f2}) if one wants to use only the
dominant term, and $\Gamma \partial_\theta H_1$ is given by
Eq.~(\ref{eqn:GaD}). The result in Sec.~\ref{sec:2} ensures
that the controlled Hamiltonian $H_c$ with the control term $f$
given by Eq.~(\ref{eqn:ctl}) has an invariant surface whose
equation is given by Eq.~(\ref{eqn:surf}). These expressions
provide the basis to investigate the control of the stochastic
transport governed by magnetic perturbations in fusion devices in
the two cases, core tearing modes and boundary stochastic layers
as done in Section $4$.

\section{Control of stochastic transport in fusion devices}
\label{sec:4}
\subsection{Control of the loss of confinement governed by coupled tearing modes}
\label{sec:41}
\begin{figure}
\unitlength 1cm
\begin{picture}(15,6.3)(0,0)
\put(0,0){\epsfig{file=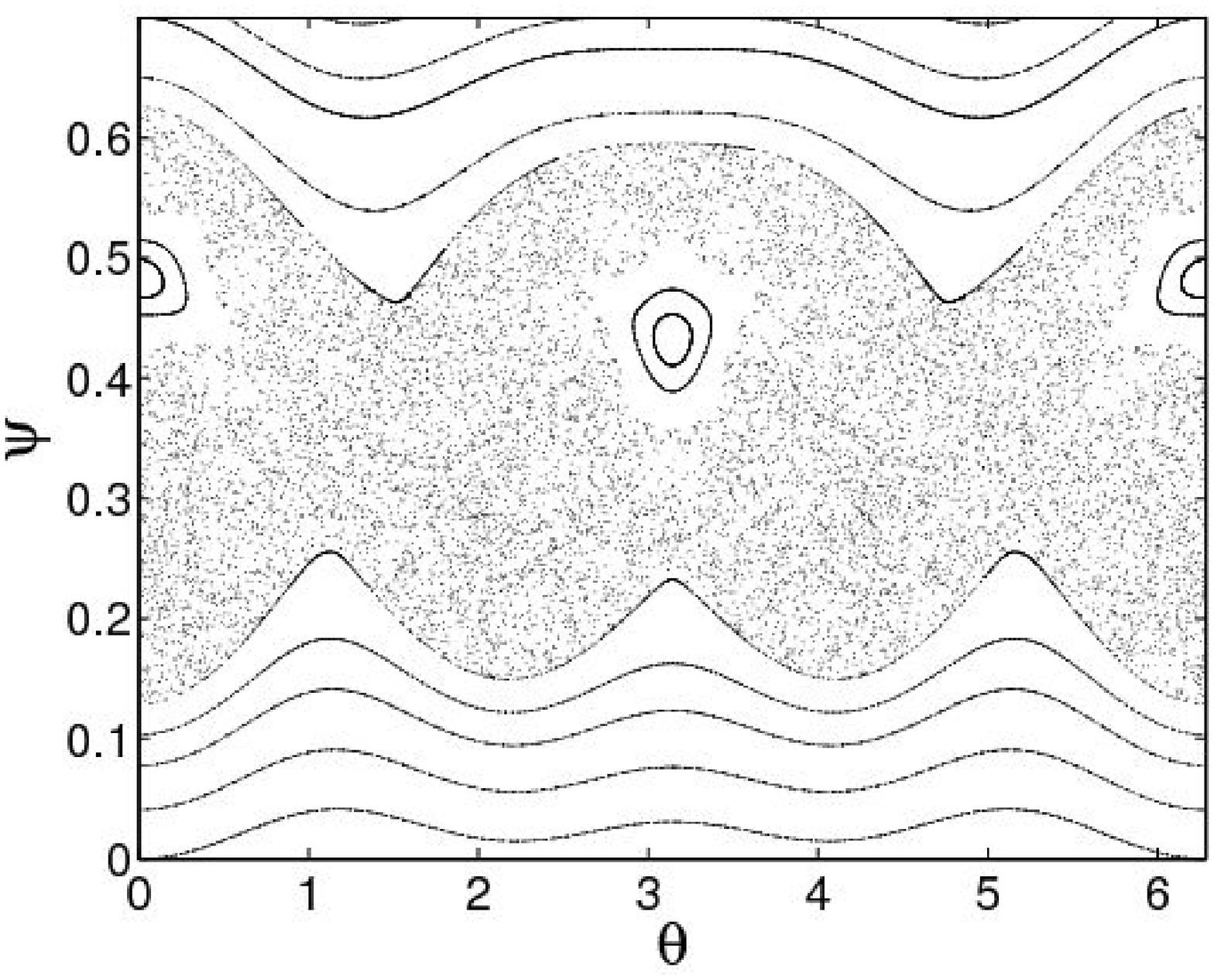,width=7.5cm,height=6.3cm}}
\put(8,0){\epsfig{file=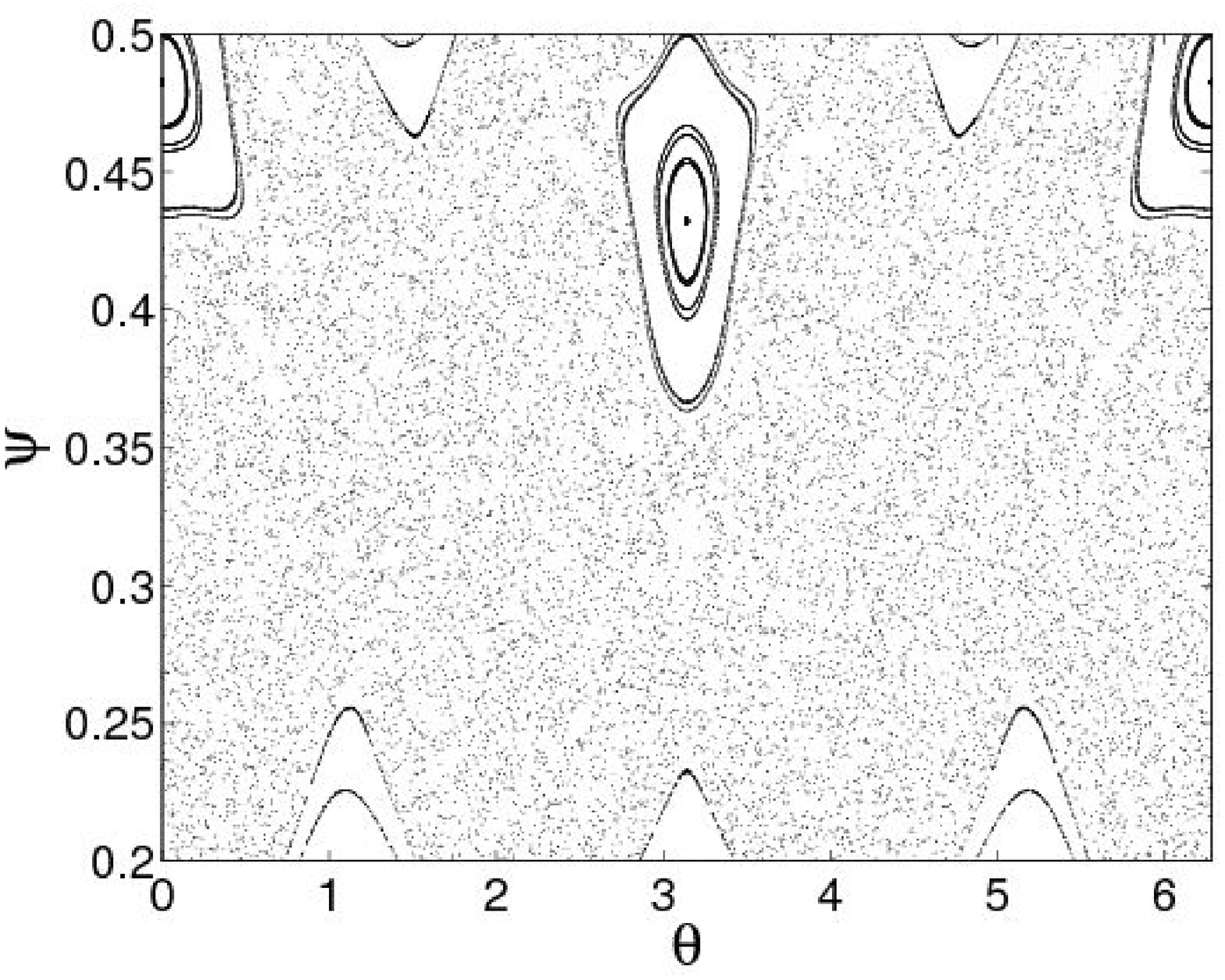,width=7.5cm,height=6.3cm}}
\end{picture}
\caption{Poincar\'e sections of Hamiltonian (\ref{eqn:HA}) with $H_1$ given by Eq.~(\ref{eqn:H11}) with $\varepsilon=0.004$ for $\psi\in[0,0.7]$ (left panel) and for $\psi\in[0.2,0.5]$ (right panel).}
\label{fig:1}
\end{figure}

We consider a Hamiltonian system~(\ref{eqn:Hms}) where $\psi$ is
the normalized toroidal flux, $\psi=1$ at the plasma boundary,
with a $q$-profile as chosen in Refs.~\cite{bale98a,misg03,abdu04}
\begin{equation}
\label{eqn:qprof}
q(\psi)=\frac{4}{(2-\psi)(2-2\psi+\psi^2)}.
\end{equation}
In this expression, $q(\psi)$ is a monotonic function of $\psi$
that varies between $1$ on axis, for $\psi=0$, and $4$ at the
plasma edge, $\psi=1$. With such an expression, the slope $d
\log(q) / d\psi$ is approximately constant and equal to
$\log(q(\psi=1))$. The results do not depend on the precise form of
the safety factor profile and can be readily extended to any
profile. We are interested here in the case of two
tearing modes with low rational numbers, in practice $(m,n)=(3,2)$
and $(m,n)=(2,1)$ where $m$ is the poloidal mode number and $n$
the toroidal mode number, locked to one another. In the case
analyzed here both perturbations have equal magnitude
characterized by $\varepsilon$. The latter parameter is assumed to
be small, but such that the onset to stochastic transport is
reached with a magnetic perturbation of the form
\begin{equation}
\label{eqn:H11}
H_1(\psi,\theta,\varphi)=
\cos(2\theta-\varphi)+\cos(3\theta-2\varphi),
\end{equation}
so that the Hamiltonian of the system is
\begin{equation}
\label{eqn:HA}
    H=\int \frac{d\psi}{q(\psi)}+\varepsilon
    H_1(\psi,\theta,\varphi).
\end{equation}
The resonant surfaces are found to be located on $\psi_{3,2} \approx
0.266$ ($q(\psi_{3,2})=3/2$) and $\psi_{2,1} \approx 0.456$
($q(\psi_{2,1})=2$). Expanding to second order the Hamiltonian $H$
in the vicinity of the two resonant surfaces and retaining the
resonant term of $H_1$, one recovers the characteristic
Hamiltonian of the pendulum~\cite{ghendrih_1992} that allows one
to define the so-called unperturbed island width $\delta$ in units of
$\psi$~:
$$
    \delta= 2 \left(\frac{\varepsilon
    }{\frac{-d(1/q(\psi))}{d\psi}}\right)^{1/2}.
$$
A Poincar\'e section of the dynamics given by Eqs.~(\ref{eqn:HA})-(\ref{eqn:H11}) is represented on
Fig.~\ref{fig:1} for $\varepsilon=0.004$. For this value of
$\varepsilon$, the island widths are $\delta_{3,2}\approx 0.125$ and
$\delta_{2,1}\approx 0.147$ so that a Chirikov parameter
$\sigma=(\delta_{2,1}+\delta_{3,2})/(\psi_{2,1}-\psi_{3,2})$ can
be computed and is about 1.4, hence larger than the
reference value for the onset of large scale transport between the
resonant surfaces $\psi_{3,2}$ and $\psi_{2,1}$ as readily
observed on Fig.~\ref{fig:1}. However, $\varepsilon$ is small
enough that significant remnant islands are still present.

The control term we apply is given by Eq.~(\ref{eqn:f2}) where
\begin{eqnarray*}
    && {\partial_\psi H_1}(\psi_0,\theta,\varphi)=0,\\
    && \Gamma\partial_\theta H_1(\psi_0,\theta,\varphi)= \frac{2}{2\omega-1}\cos(2\theta-\varphi)+\frac{3}{3\omega-2}\cos(3\theta-2\varphi).
\end{eqnarray*}
For $\psi_0$, we choose $\psi_0= 0.35$ hence between the two
resonant surfaces $\psi_{3,2}\approx 0.266$ and $\psi_{2,1}\approx 0.456$.
Choosing other values of $\psi_0$ is equivalent to moving the
barrier one wants to create. The expression of the partial control
term $f_2$ is given by Eq.~(\ref{eqn:f2})~:
\begin{equation}
\label{eqn:f2tearing}
f_2(\theta,\varphi)=\Big(-\frac{d(1/q(\psi))}{d\psi}\Big)\Big|_{\psi=\psi_0}
\left(\frac{2\cos(2\theta-\varphi)}{2\omega-1}+\frac{3\cos(3\theta-2\varphi)}{3\omega-2}
\right)^2,
\end{equation}
where
$\omega=1/q(\psi_0)=(2-\psi_0)(2-2\psi_0+\psi_0^2)/4$ and where we used the fact that $H_1$ does not depend on $\psi$. For the
present value of $\psi_0$ one finds $q(\psi_0)\approx 1.7$ and thus
$\omega\approx 0.587$. The full control term $f$ creates an invariant
torus whose location is given by
\begin{equation}
\label{eqn:MS3tor}
\psi=\psi_0-\varepsilon \left(\frac{2}{2\omega-1}\cos(2\theta-\varphi)+\frac{3}{3\omega-2}\cos(3\theta-2\varphi) \right).
\end{equation}
The magnitude of the angle modulation of the invariant torus
labeled by $\psi$ is relatively large (larger than
$40\varepsilon$).

\begin{figure}
\unitlength 1cm
\begin{picture}(15,6.3)(0,0)
\put(0,0){\epsfig{file=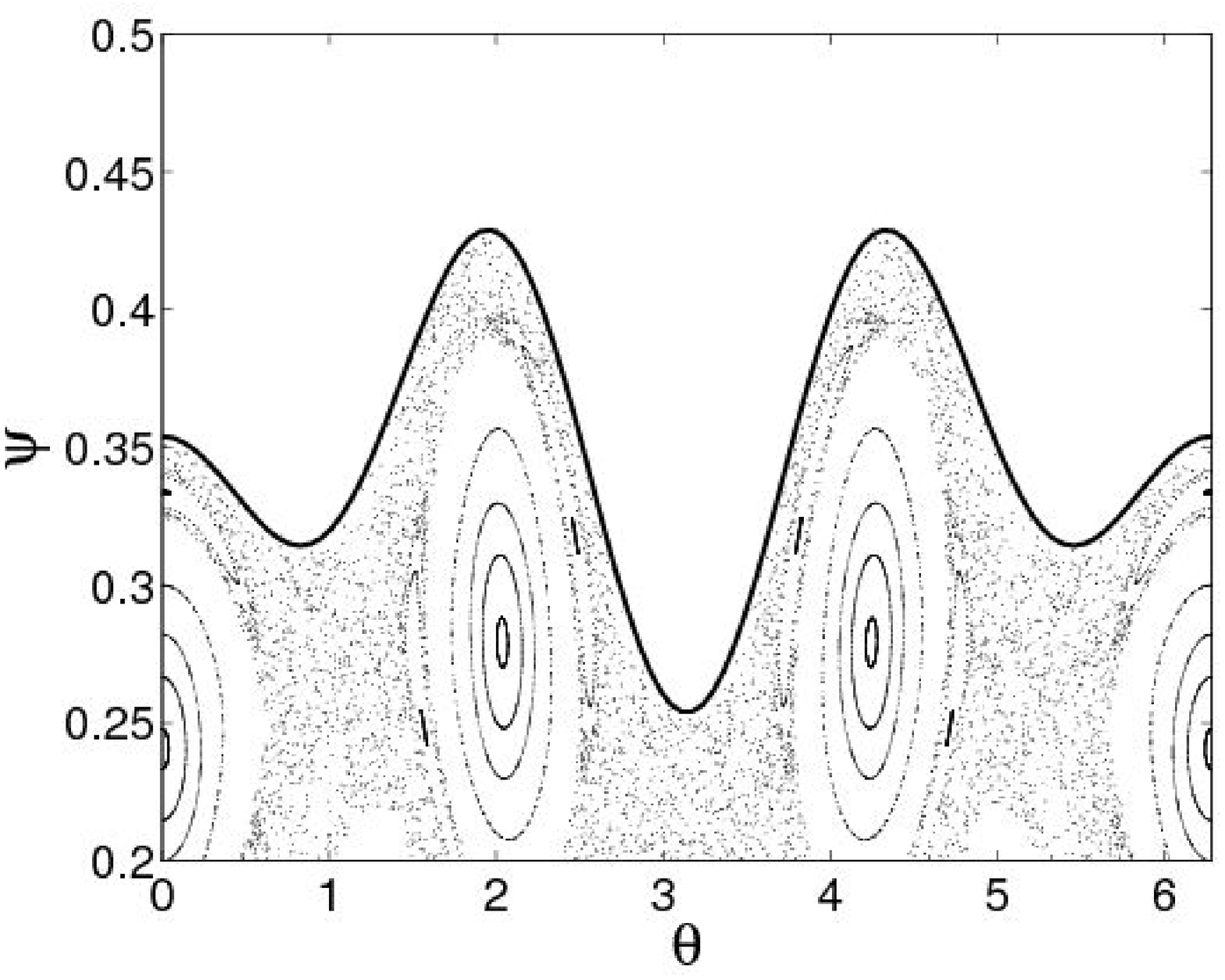,width=7.5cm,height=6.3cm}}
\put(8,0){\epsfig{file=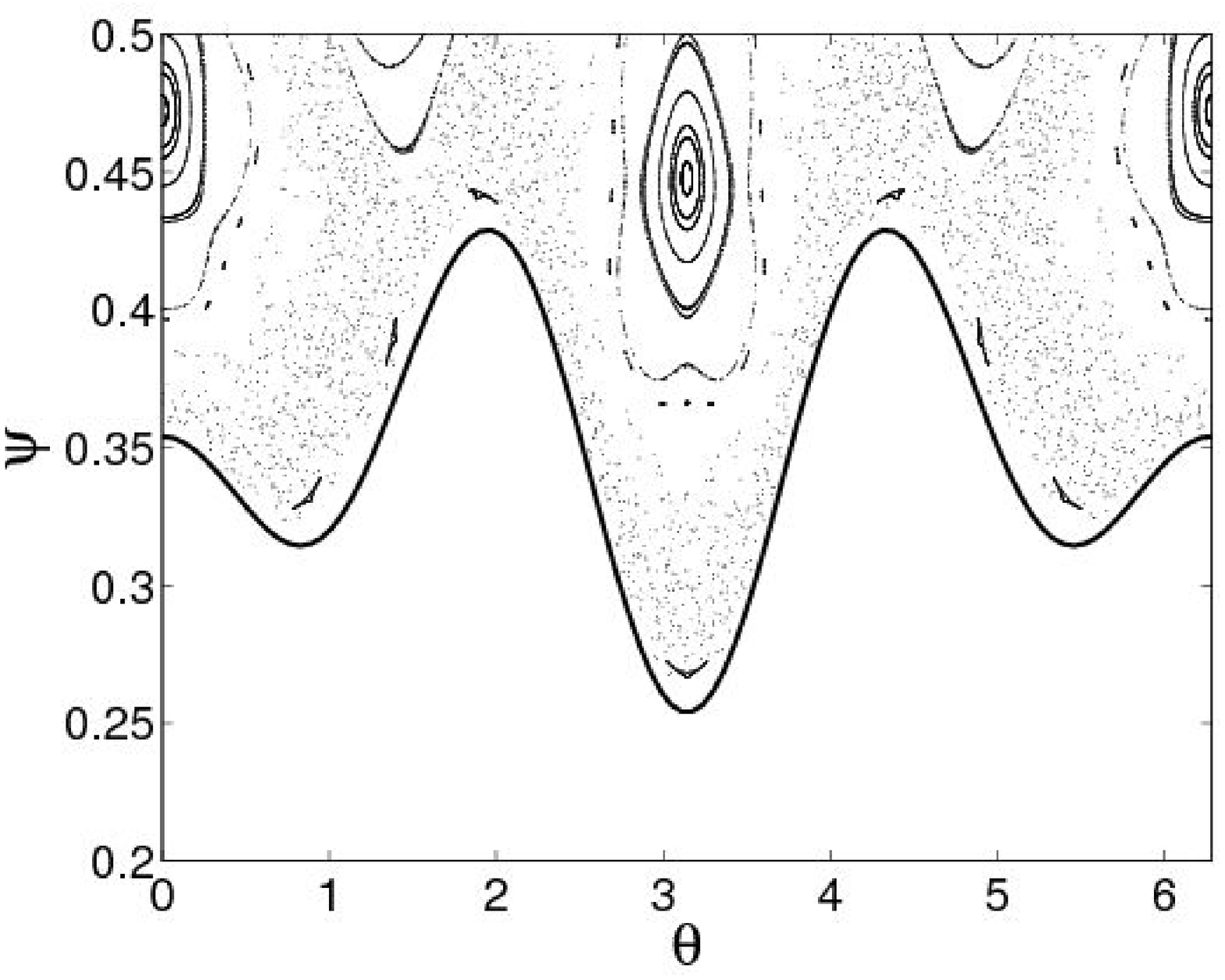,width=7.5cm,height=6.3cm}}
\end{picture}
\caption{Poincar\'e sections of Hamiltonian
(\ref{eqn:HA}) with the control term $f_2$ given by
Eq.~(\ref{eqn:f2tearing}) with $\varepsilon=0.004$ and $\psi_0=0.35$
using $\Omega=1$, thus with no localization of the control term~:
with initial conditions below (left panel) or above (right panel) the surface
given by Eq.~(\ref{eqn:MS3tor}) plotted in bold.}
\label{fig:2}
\end{figure}

For the localization function $\Omega$, we use two choices~: The
first one is $\Omega(|x|)=1$ for all $|x|\in{\mathbb R}^+$, i.e.
without localization. Such a control procedure appears to be more readily applicable in fusion plasmas where it is difficult to device an electromagnetic perturbation localized in a narrow region of the core plasma. The Poincar\'e sections of trajectories of
the controlled Hamiltonian $H_0+\varepsilon H_1+\varepsilon^2 f_2$
are represented in Fig.~\ref{fig:2} for
trajectories started from below or from above the invariant torus
given by Eq.~(\ref{eqn:MS3tor}) and represented by the bold curve. We notice in particular that the
trajectories started from below (resp.\ above) the invariant torus
remain below (resp. above) it. The impact of the control term is
noticeable away from the target torus $\psi_0$ since one can
observe much larger remnant islands in the vicinity of the two
resonant surfaces compared to the case without control.

\begin{figure}
\unitlength 1cm
\begin{picture}(15,6.3)(0,0)
\put(0,0){\epsfig{file=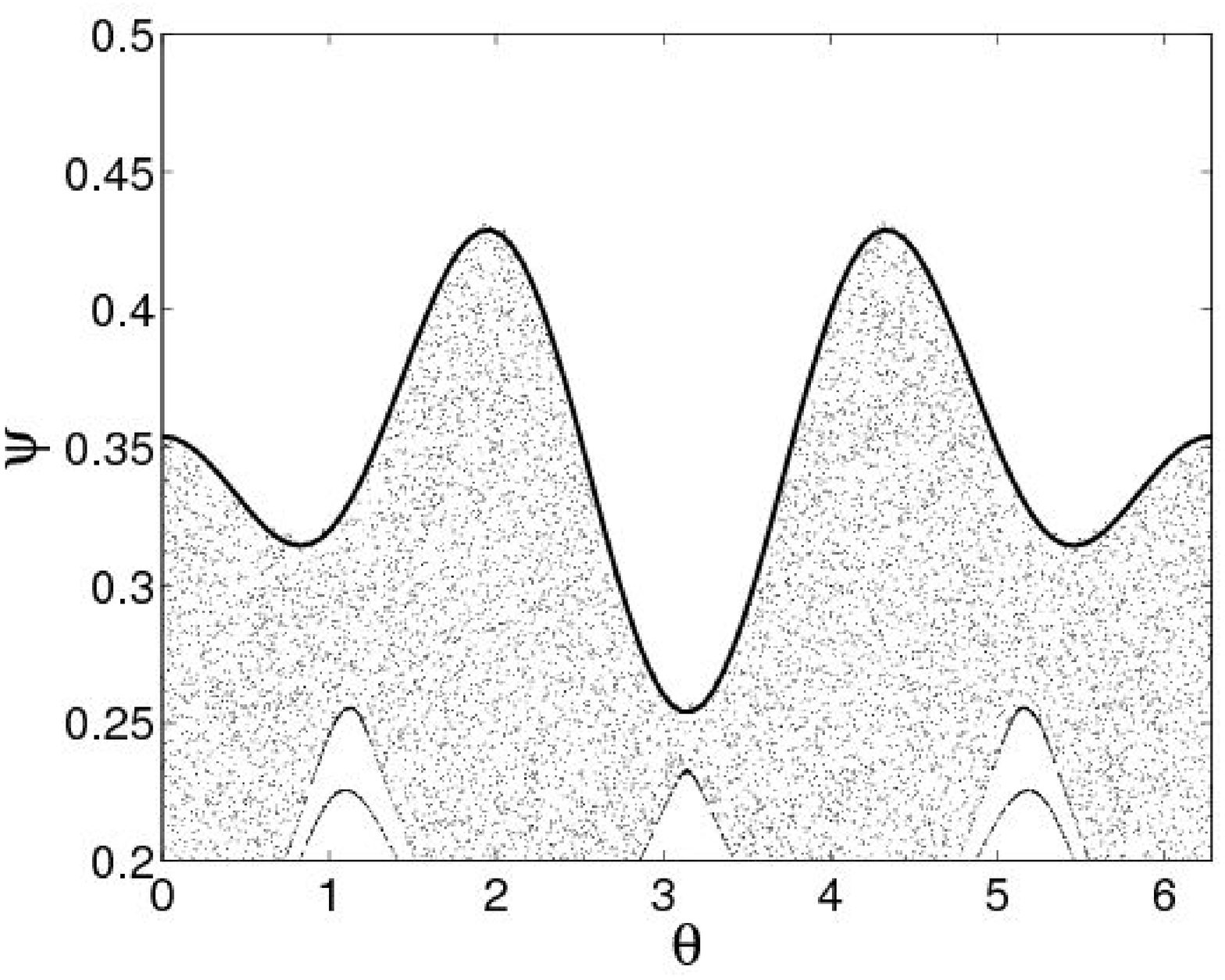,width=7.5cm,height=6.3cm}}
\put(8,0){\epsfig{file=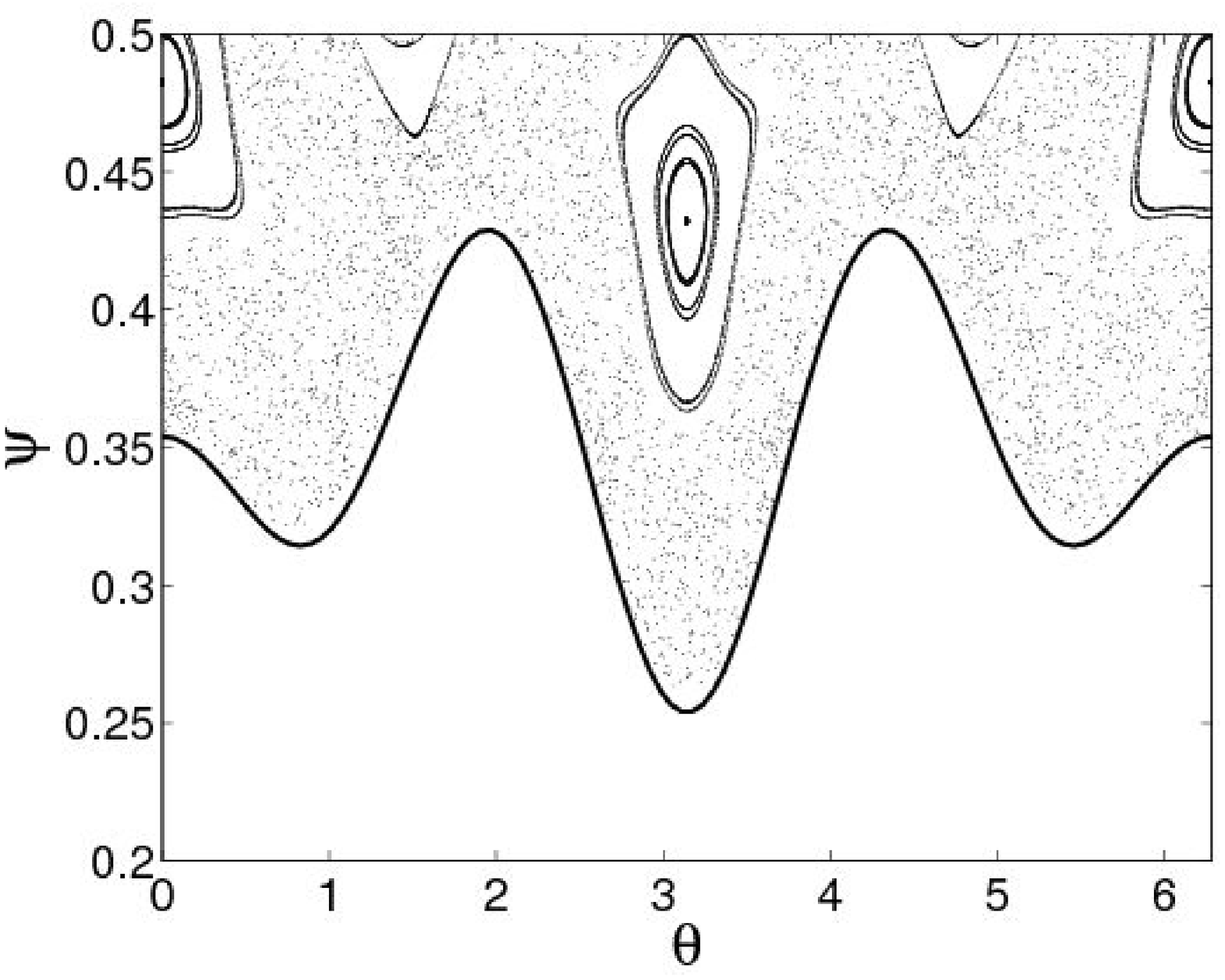,width=7.5cm,height=6.3cm}}
\end{picture}
\caption{Poincar\'e sections of Hamiltonian (\ref{eqn:HA}) with the control term $f_2$ given by Eq.~(\ref{eqn:f2tearing}) with $\varepsilon=0.004$ and $\psi_0=0.35$ using $\Omega=\Omega_{loc}$~: with initial conditions below (left panel) or above (right panel) the surface given by Eq.~(\ref{eqn:MS3tor}) plotted in bold.}
\label{fig:3}
\end{figure}

The second choice of $\Omega$ is a localization function. Although such a localized control term has yet no clear application to fusion plasmas, it demonstrates that the efficiency of the control procedure is not tied to the perturbation of the whole phase portrait but only to a specific region of the phase space defined by Eq.~(\ref{eqn:MS3tor}). It is therefore illustrative to consider such a control procedure in order to check in particular that the leading effect of the control scheme is restoring a selected magnetic surface and not a wide range cancellation of the effects of the perturbation. We
choose $\Omega_{loc}(|\delta\psi|)=1$ for $|\delta\psi| \leq
\delta\psi_\alpha$, $\Omega_{loc}(|\delta\psi|)=0$ for
$|\delta\psi| \geq \delta\psi_\beta$ and a third order polynomial
for $|\delta\psi| \in ]\delta\psi_\alpha,\delta\psi_\beta[$ for
which $\Omega_{loc}$ is a $C^1$-function, i.e.\
$\Omega_{loc}(|\delta\psi|)=1-(|\delta\psi|-\delta\psi_\alpha)^2
(3\delta\psi_\beta-\delta\psi_\alpha-2|\delta\psi|)/(\delta\psi_\beta-\delta\psi_\alpha)^3$.
In principle, one can choose arbitrarily small values for
$\delta\psi_\alpha$ and $\delta\psi_\beta$ if one uses the full control term $f$ given by
Eq.~(\ref{eqn:ctl}). However, since $f$ is given by a series, it
is more practical to consider the truncated control term $f_2$ (or
a truncation of the series which gives the control term $f$). Then
the value of $\delta\psi_\alpha$ has to be not too small such that
the set $\{({\bf A},{\bm\theta}) \mbox{ s.t.}\Omega_{loc}({\bf
A},{\bm\theta}) =1\}$ contains the invariant torus which, for
$f_2$, is $\varepsilon^3$-close to the one obtained using the
complete control term. This leads to a restriction for
$\delta\psi_\alpha$, $\delta\psi_\alpha \gtrsim 10^{-3}$ to embed the
lowest order in the control term expansion. For the numerics, we
choose $\delta\psi_\alpha=0.01$ and $\delta\psi_\beta=0.02$. A
Poincar\'e section of the dynamics of $H_c$ given by
Eq.~(\ref{eqn:HmsC}) with the control term $f_2$ is represented on
Fig.~\ref{fig:3} for $\varepsilon=0.004$ and for $\Omega=\Omega_{loc}$. The bold curve
corresponds to the invariant torus given by Eq.~(\ref{eqn:MS3tor})
that has been created by the addition of the control term which is
localized around this surface.

From these figures, we clearly see that the upper and lower parts of phase space are very similar to the ones of Fig.~\ref{fig:1} (without control). More precisely, we notice that the structure of the resonant islands is not modified, even the neighboring ones. What has changed is the dynamics in the neighborhood of the bold curve due to the action of the localized control. There is now an invariant torus which prevents the motion to diffuse from the lower part to the upper part of phase space. These two parts are invariant by the dynamics of the controlled Hamiltonian.\\
The invariant torus created by the localized control persists to higher values of $\varepsilon$. For $\varepsilon \geq 0.1$, the trajectories start to diffuse through the broken invariant surface. The diffusion can be reduced or even suppressed by taking into account higher order terms $f_s$ for $s\geq 3$ in the control term series. We point out that the value of $\varepsilon$ for which the partial control term $f_2$ is efficient depends on the choice of $\psi_0$. There is freedom to choose the initial surface $\psi_0$ provided that $q(\psi_0)$ is sufficiently irrational.

Let us define the norm of a function
$f(\theta,\varphi)=\sum_{m,n}f_{mn}\cos (m\theta-n\varphi)$ as
$\Vert f\Vert=\max_{m,n}\vert f_{mn}\vert$. If we compare the
relative size of the control we obtain~:
$$
\frac{\Vert \varepsilon^2  f_2 \Vert }{\Vert \varepsilon H_1 \Vert }\approx 35 \varepsilon,
$$
typically 14\% of $\varepsilon H_1$ for $\varepsilon=0.004$. The
magnitude of the control term computed in such a way appears to be
a small fraction of the magnitude of the initial perturbation that
led to the stochastic transport. Moreover, when considering the localized control term, the control only acts
on a finite and small portion of the phase space
$[0,1]\times{\mathbb T}^2$ around the invariant surface (the size of the support of $\Omega_{loc}$ is 4\%).

Among the aims of a control scheme, that of restoring a region with closed magnetic surfaces in a disruptive phase can prove to be very valuable even if all nested magnetic surfaces are not recovered. Indeed after the thermal quench, when most of the plasma kinetic energy has been lost, the standard disruption scenario enters the so-called current quench. In this phase the plasma current strongly decreases which leads to a loop-voltage spike in order to sustain the plasma current. Such a voltage spike creates conditions for the transfer of the energy from the poloidal system to the plasma and specifically for the generation of runaway electrons. These are considered as a major problem in a device like ITER since these runaway electrons will lead to very strong energy deposition on reduced areas and deep into the material given the slowing down distance of these energetic particles. As a consequence, this phase of the disruption can lead to serious damage to wall components. Defining appropriate control tools to reduce the energy transfer to a runaway population or to lower the runaway population is thus a key task for ITER. 
	One can readily assume that already in the thermal quench phase, a broad spectrum of magnetic perturbations is responsible for the confinement loss due to the onset of large scale transport governed by the overlap of the various modes of the spectrum. Among these modes, one can expect that the low $(m, n)$ modes, such as the $(3, 2)$ and $(2, 1)$, to overlap and govern a strongly enhanced transport as required for the thermal quench. When applying the control magnetic perturbation towards before the onset of the current quench, very likely when the stochastic core region comes into contact with the walls, one aims at restoring a magnetic surface between the two resonant surfaces.  In practice, this yields a transport barrier (with respect to the very degraded confinement during the thermal quench) that will separate the core region, where the total current will be maintained, from the edge region where all the current will be lost due to the connection to the wall components. Such an insulation ought to prevent the current quench and the generation of runaway electrons. Should the loop-voltage spike be only partially suppressed, and a fraction of runaway electrons still generated, these should remain confined within the transport barrier. The control scheme applied to a tokamak disruptive phase thus appear to provide conditions for a soft landing with no current quench.
	Applicability of such a scheme depends of course on the possibility to generate the spectrum of the control term. Let us consider the spectrum of $f_2$ as given by Eq.~(\ref{eqn:f2tearing}). It consists of four modes, the two modes corresponding to the tearing modes that define the region where a magnetic surface is to be restored, hence the $(3,2)$ and $(2,1)$ modes and their nonlinear combination, namely the $(1,1)$ and $(5,3)$ modes. The relative magnitude of the modes determines the value of $\omega$ and therefore the location of the barrier in terms of the safety factor $q$. For a given discharge, this parameter can be preset so that the activation of the control scheme will not require a real time computation of the required spectrum. The means to generate such a magnetic perturbation is to implement dedicated coils at the plasma boundary. Of course, this means that one must contemplate the non-localized control. Although it is not the purpose of the present paper to design the appropriate coils to generate these modes, one can underline the main features of the coil system. First, the required coils are significantly simpler than the coils required for boundary plasma control either of the ergodic divertor kind~\cite{jaku04,ghendrih_1996} or to control ELMs~\cite{stella2}. This is due to the low poloidal mode number of the modes that will therefore exhibit a moderate radial decay. Furthermore, the magnitude of the control term is typically 14\% lower than required to generate stochasticity as required for boundary plasma control. Of course a proper project would have to address many issues left open here, like the effect of a mismatch in magnitude of the control term, or the effect of spurious modes on the overall control efficiency. However, the tests that have been performed to examine the robustness of the control procedure (see Refs.~\cite{guido1,guido2}) give us confidence that these effects should only lead to moderate reduction of the control efficiency. These favourable features show that the application of the control scheme to prevent the current quench during a disruption should not face outstanding design difficulties especially when compared to the consequences of uncontrolled runaway generation.

\subsection{Control of stochastic boundary plasmas}
\label{42}
For the second example, we consider a magnetic perturbation which
models the magnetic field lines in an ergodic
divertor~\cite{ghendrih_1996,abdu98,abdu99,jaku04,abdu04}. We use
the $q$-profile given by Eq.~(\ref{eqn:qprof}) and the following
magnetic perturbation
\begin{equation}
\label{eqn:MS2}
    H_1(\psi,\theta,\varphi)=\sum_m (-1)^m \frac{\sin[(m-m_0)\theta_d]}{\pi(m-m_0)}\psi^{\bar{m}/2} \cos(m\theta-n\varphi),
\end{equation}
where $n=2$, $\theta_d=\pi/3$, $m_0=6$ and the sum ranges from
$m_0-4$ to $m_0+4$. We have chosen $\bar{m}=m$. In experiments, the strong modulation of the
toroidal magnetic field between the inner region and the outer
region of the torus can lead to a significant departure from this
approximation. This is the case for the Tore Supra experiment with $\bar{m}\sim m/2$ and the Textor DED experiment with $\bar{m}\sim 2m$~\cite{ghendrih_1994}. However, such aspects of the perturbation, while
important for the engineering constraints, do not modify
significantly the computation of the control term. Also, the
values chosen for the mode numbers are relevant for the Textor DED,
a rather low $m$ and $n$ configuration that is required to provide a
reasonable stochastic boundary in spite of the DED coil location
on the high field side~\cite{ghendrih_1994}. A possible operating regime of the DED is to induce a rotation of the magnetic perturbation at constant frequency for all the modes of the spectrum by an appropriate phasing of the coils. The control scheme derived here does not depend on the choice of the rotation frequency provided the magnetic perturbation of the control term is designed to achieve the same rotation frequency as the main magnetic perturbation. For the purpose of the present analysis, the rotation frequency is not taken into account, and we shall concentrate on the control of a stochastic boundary induced by a steady state perturbation. 

Few modes are then
effective in the spectrum given by Eq.~(\ref{eqn:MS2}) and the largest extent
of the perturbed edge region is achieved when the main mode of the
spectrum, $m_0=6$ and $n=2$ (hence $q(\psi_{6,2})=3$) is located
within the edge region, at $\psi_{6,2}\approx 0.75$ for $q_{edge}=4$
for the chosen profile of the safety factor. The region chosen
for the control, $\psi_0\approx 0.9$, $q(\psi_0)\approx 3.6$ thus appears
to be located in the vicinity of the mode $(7,2)$ of the spectrum
characterized by a large amplitude ($\sim 0.8$ to be compared to
$1$ for the main component $(6,2)$).

\begin{figure}
\unitlength 1cm
\begin{picture}(15,6.3)(0,0)
\put(0,0){\epsfig{file=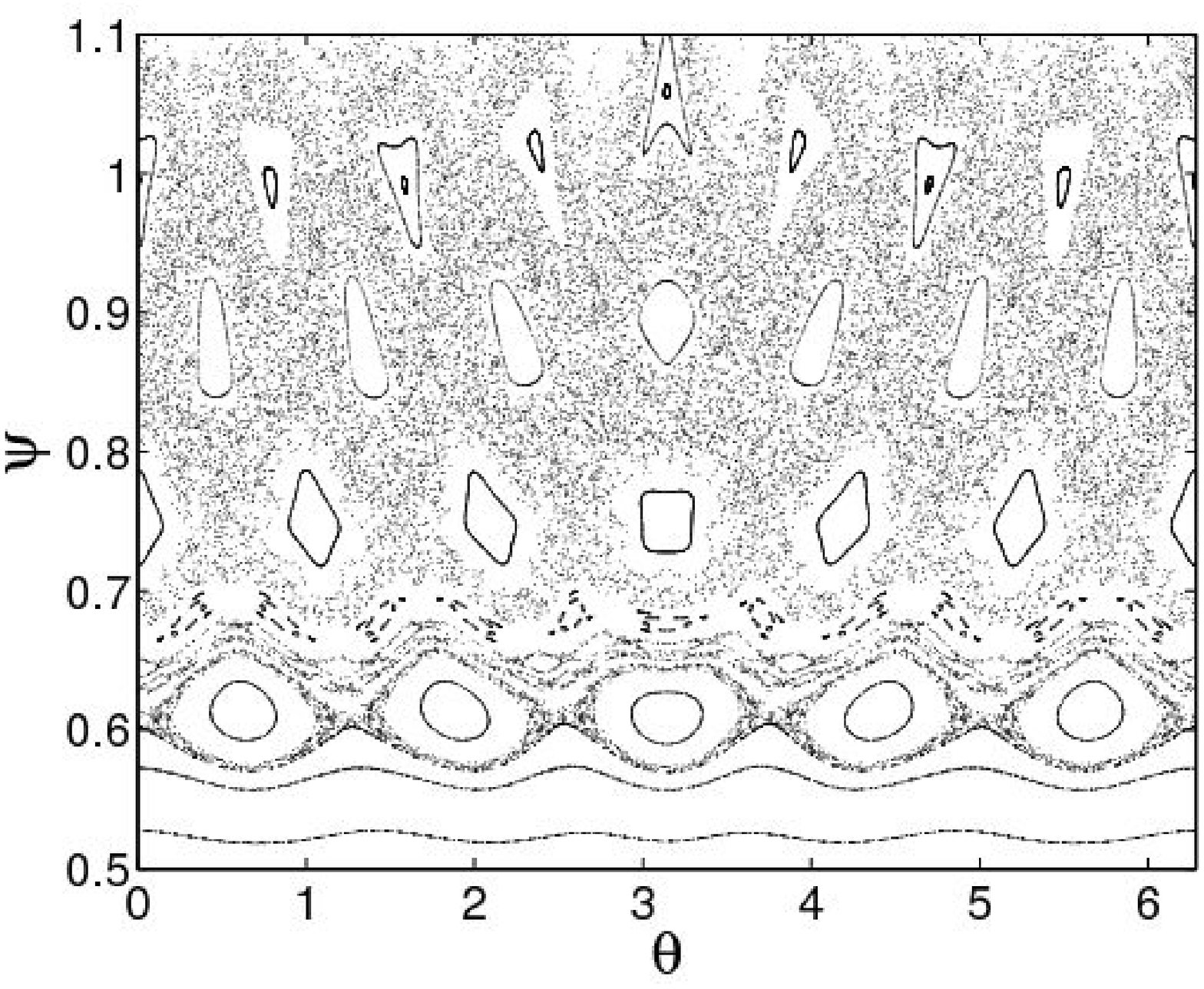,width=7.5cm,height=6.3cm}}
\put(8,0){\epsfig{file=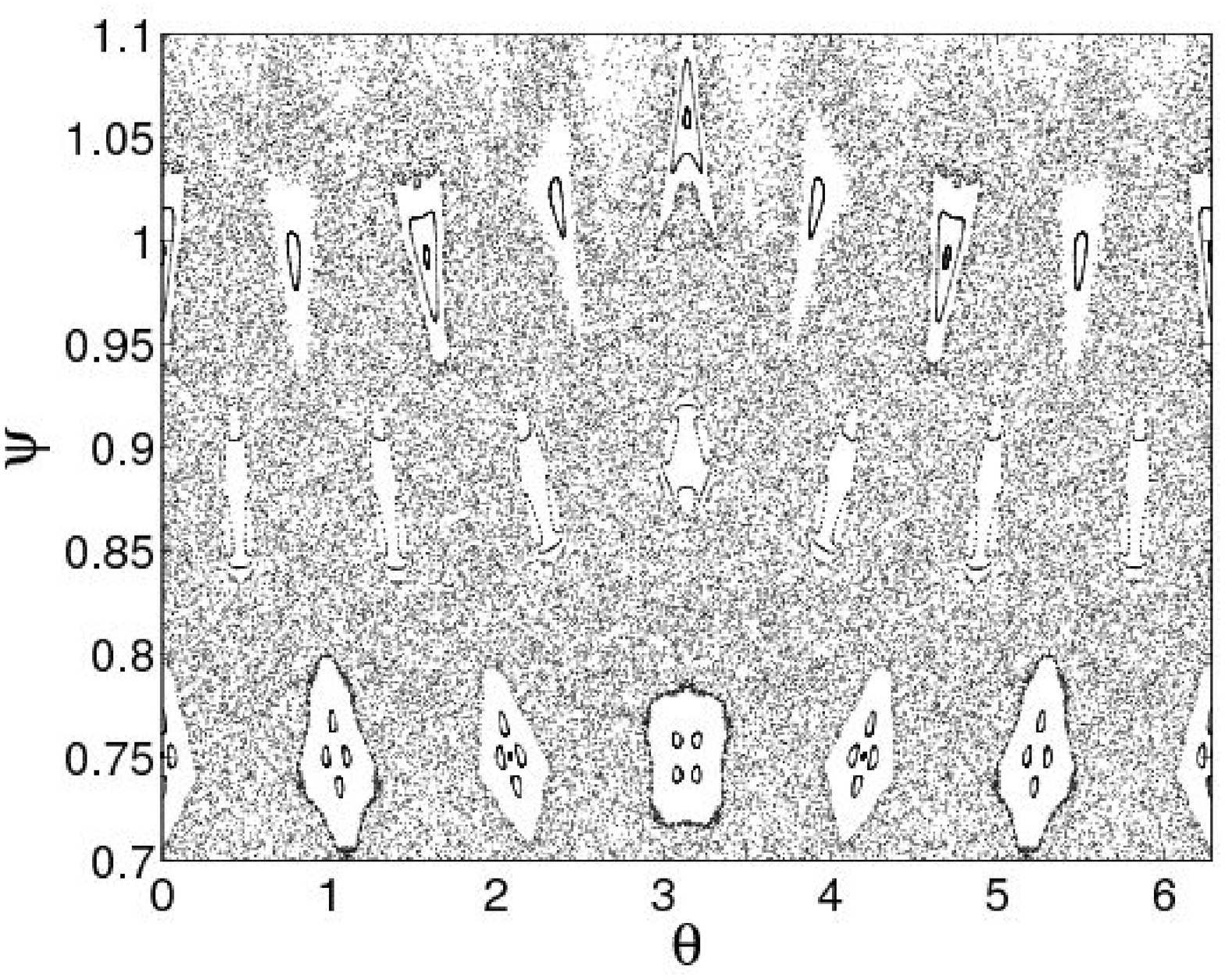,width=7.5cm,height=6.3cm}}
\end{picture}
\caption{Poincar\'e sections of $H$ given by Eq.~(\ref{eqn:HA}) with $H_1$ given by Eq.~(\ref{eqn:MS2}) with $\varepsilon=0.003$ for $\psi\in [0.5,1.1]$ (left panel) and for $\psi \in [0.7 1.1]$ (right panel).}
\label{fig:4}
\end{figure}

One well documented operational limit of stochastic boundary
plasmas is achieved when the stochastic region extends over a too
large radial extent \cite{zabiego_2000, ghendrih_2000b, ghendrih_2000a} and when the stochastic boundary actually reaches
the wall, namely when all invariant tori between the stochastic
region and the wall are destroyed \cite{zabiego_2000, ghendrih_2000b,
ghendrih_2000a}. A Poincar\'e section of the dynamics is represented on
Fig.~\ref{fig:4} for $\varepsilon =0.003$ where we notice that
magnetic field lines diffuse up to outer-edge of the plasma
($\psi=1$) and to the wall.

The control term we apply is given by Eq.~(\ref{eqn:f2}) where
\begin{eqnarray*}
    && \partial_\psi H_1(\psi_0,\theta,\varphi)=\sum_m (-1)^m \frac{\sin[(m-m_0)\theta_d]}{\pi(m-m_0)}\frac{m}{2}\psi_0^{m/2-1} \cos(m\theta-n\varphi),\\
    && \Gamma\partial_\theta H_1(\psi_0,\theta,\varphi)=\sum_m (-1)^m \frac{\sin[(m-m_0)\theta_d]}{\pi(m-m_0)}\psi_0^{m/2}\frac{m}{m\omega -n} \cos(m\theta-n\varphi).
\end{eqnarray*}
The control term thus appears as the product of two terms (given above) each having a similar spectrum as the initial perturbation given in Eq.~(\ref{eqn:MS2}). As a consequence of this nonlinearity all coupled modes $m_1 + m_2$ and $n_1 + n_2$ must be generated leading in practice to a very large spectrum for the control perturbation. Since the original spectrum is in fact induced by a modulated magnetic perturbation with mode number $m_0$, $n_0$ localized in a finite poloidal arc (that govern the width of the spectrum), the control term will, at first order, have to be produced within the same poloidal arc but with a modulation $2m_0$, $2n_0$. The required coil is therefore very similar but with larger main mode numbers. The design of such a coil in the case of the DED would not be possible due to specific constraint of this design while it would have been possible, although difficult, to implement the required perturbation with the Tore Supra ergodic divertor~\cite{jaku04,ghendrih_1996}. In the case of the DED, it is thus of interest to consider the situation of a degraded control term such that only a subset of the optimum control term is implemented. Before analyzing this aspect of the control procedure and its impact on the coil design, let us consider the effect of the full control term and its impact on the transport within a stochastic boundary.

\begin{figure}
\unitlength 1cm
\begin{picture}(15,6.3)(0,0)
\put(0,0){\epsfig{file=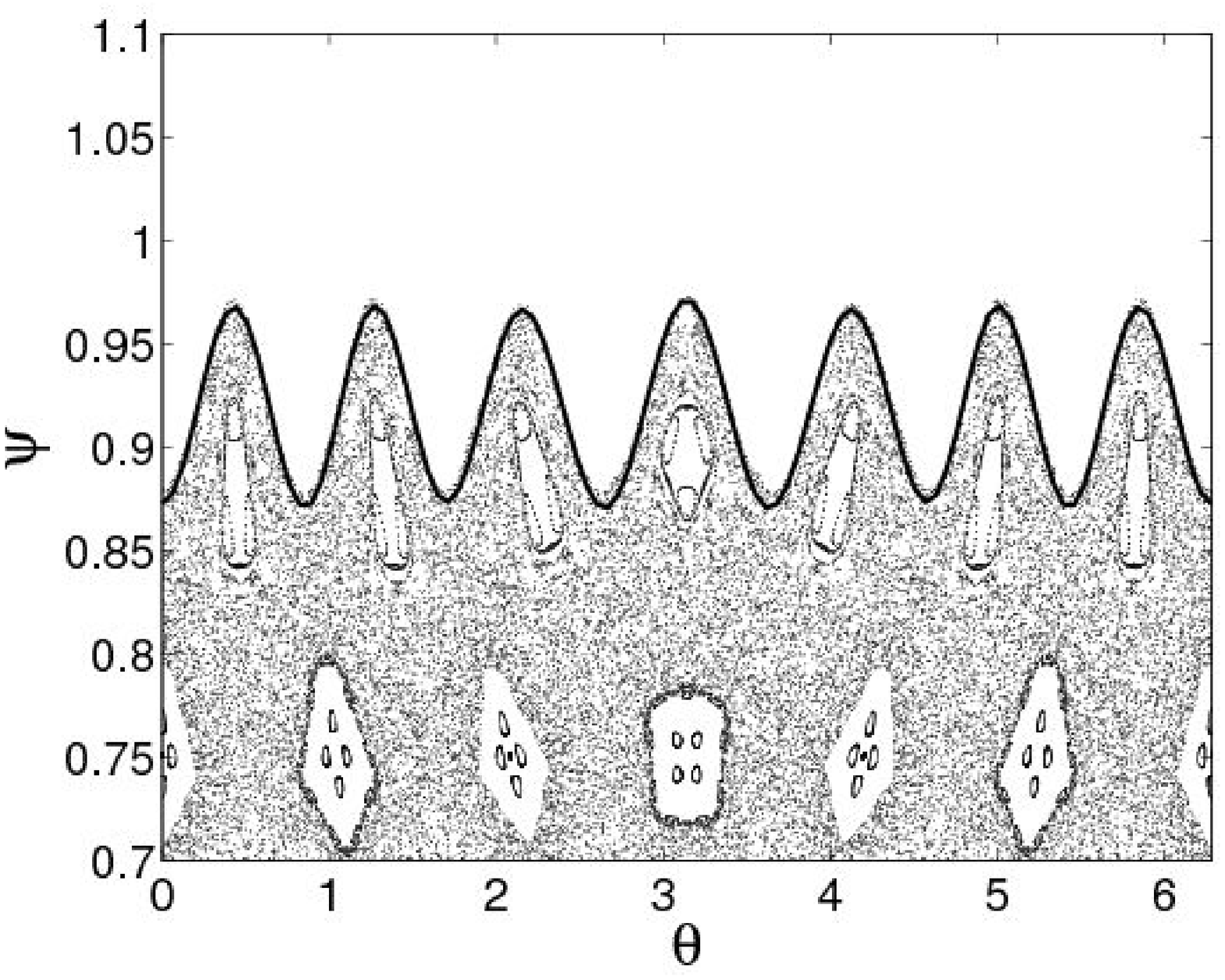,width=7.5cm,height=6.3cm}}
\put(8,0){\epsfig{file=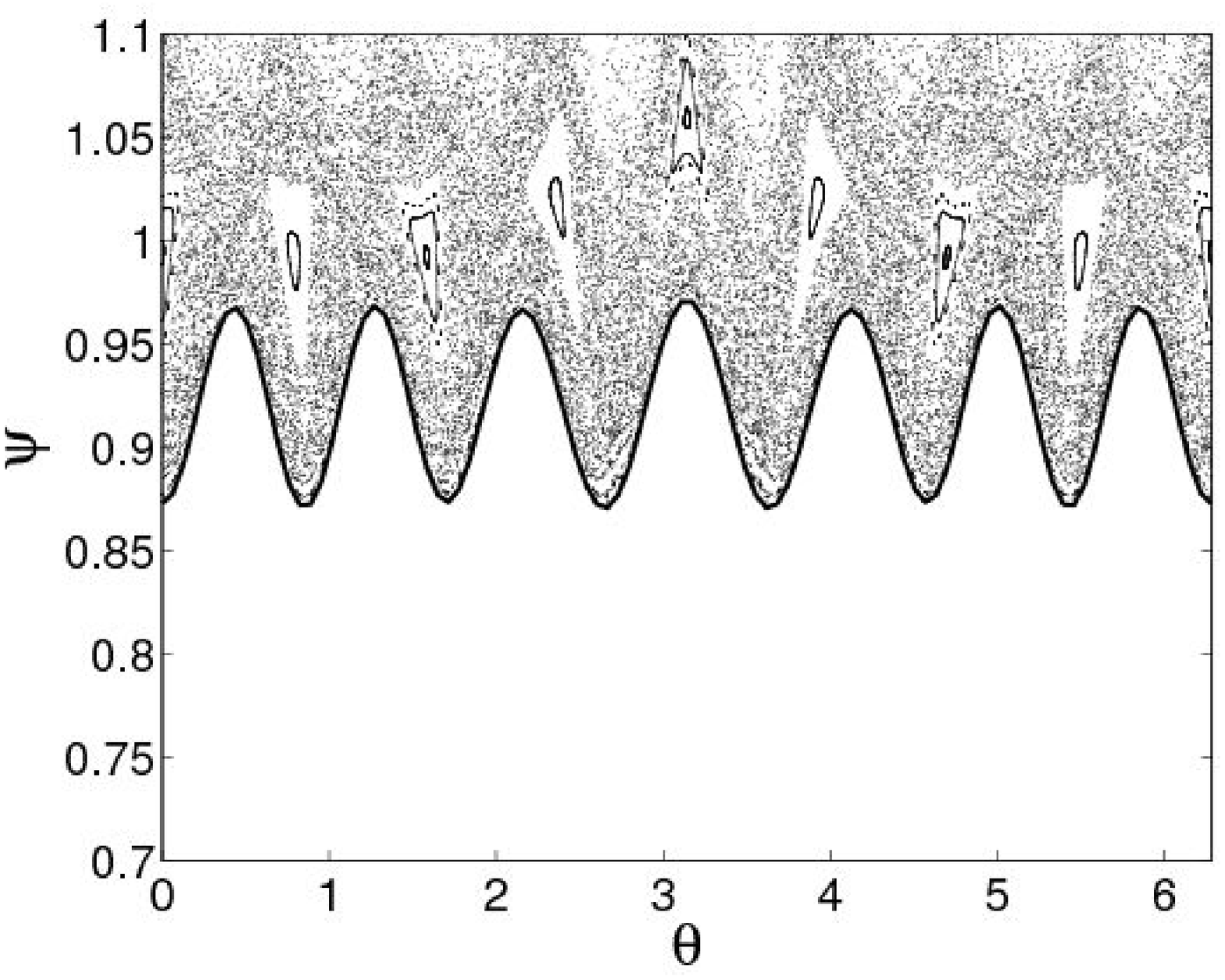,width=7.5cm,height=6.3cm}}
\end{picture}
\caption{Poincar\'e sections with $H_1$ given by Eq.~(\ref{eqn:MS2}) using $\Omega=\Omega_{loc}$, with the control
term $\varepsilon^2 f_2$ given by Eq.~(\ref{eqn:f2}) with $\varepsilon=0.003$ and $\psi_0\approx 0.92$~:
with initial conditions below (left panel) or above (right panel) the surface
given by Eq.~(\ref{eqn:surf}).}
\label{fig:5}
\end{figure}

We use the same characteristic function $\Omega_{loc}$ as in the
first example with $\delta\psi_\alpha=0.02$ and
$\delta\psi_\beta=0.03$. We choose $\psi_0\approx 0.92$. A
Poincar\'e section of Hamiltonian (\ref{eqn:HA}) with $H_1$ given by Eq.~(\ref{eqn:MS2}) and with the
control term $f_2$ given by Eq.~(\ref{eqn:f2}) for this example
with $\varepsilon=0.003$ is represented on Fig.~\ref{fig:5}.

As for the first example, we clearly see that the upper and lower
parts of phase space are very similar to the ones of
Fig.~\ref{fig:4}. In other words, this localized control does not
affect the diffusivity of magnetic lines above the magnetic
surface around $\psi=\psi_0$. However the two parts are
disconnected by the dynamics since they are invariant by the
dynamics of the controlled Hamiltonian. A magnetic surface whose
equation is given by Eq.~(\ref{eqn:surf}) has been created and
acts as a barrier to the diffusion toward the border of the
plasma. As stated above, there is experimental evidence that
indicates that such a transport barrier is sufficient to decouple
the core plasma from the edge plasma \cite{zabiego_2000, ghendrih_2000b,
ghendrih_2000a}.

The norm (as defined in the previous section) of the control $f_2$ is about 15\% of
$\varepsilon H_1$ for $\varepsilon=0.003$. Moreover, the control only acts on a
finite and small portion (around 7\%) of the phase space $[0,1]\times{\mathbb T}^2$
around the invariant surface.

\begin{figure}
\unitlength 1cm
\begin{picture}(15,6.3)(0,0)
\put(0,0){\epsfig{file=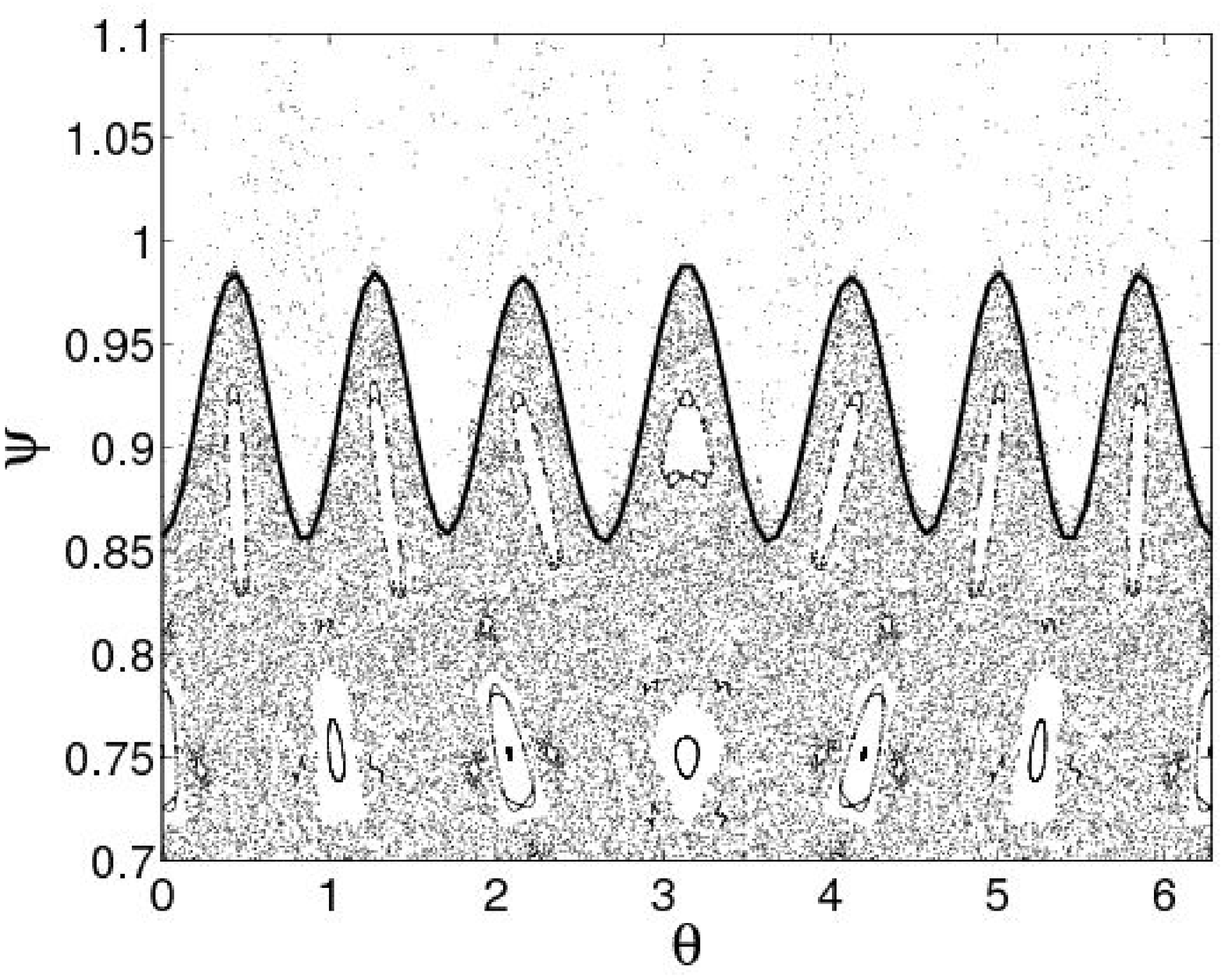,width=7.5cm,height=6.3cm}}
\put(8,0){\epsfig{file=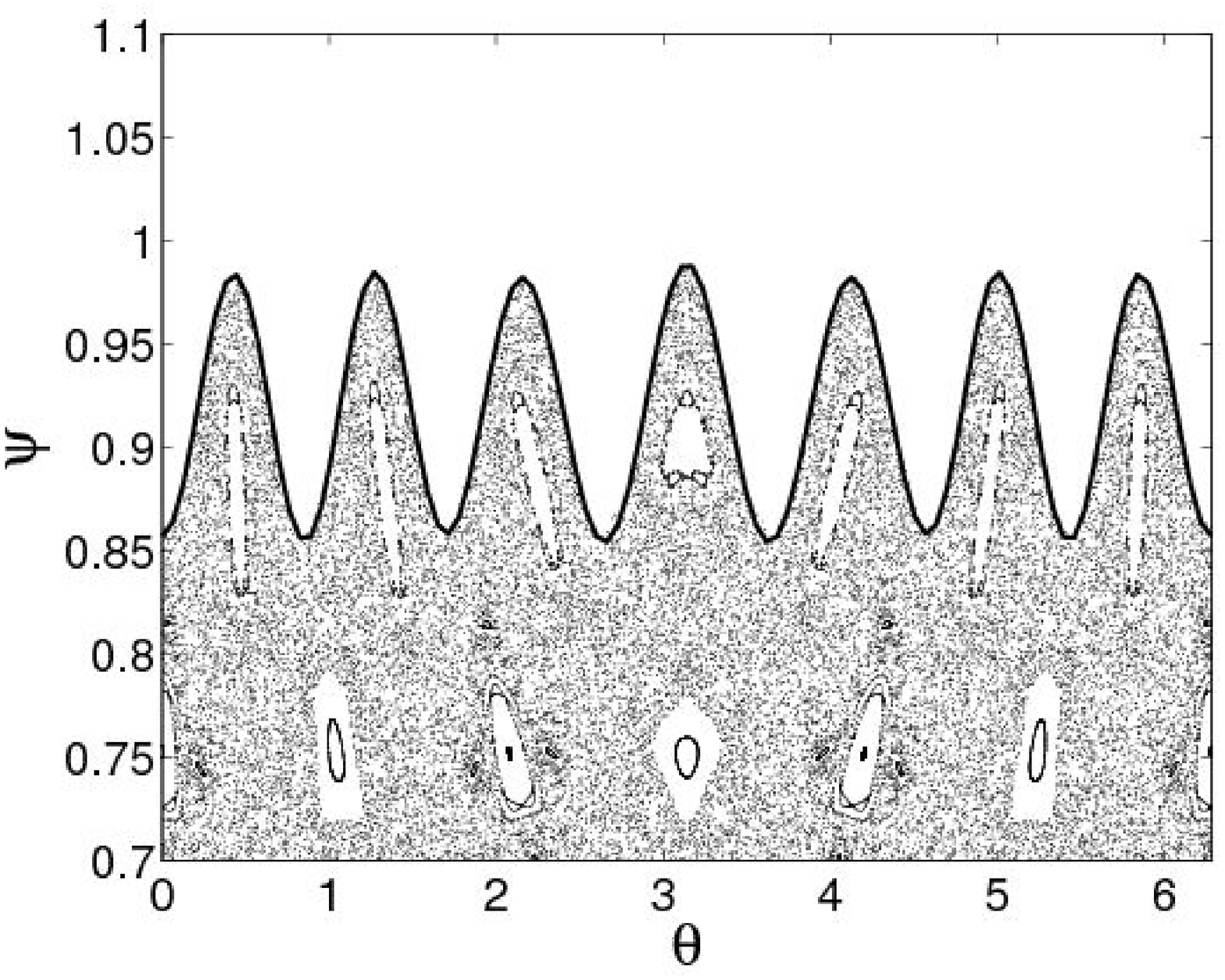,width=7.5cm,height=6.3cm}}
\end{picture}
\caption{Poincar\'e sections with $H_1$ given by Eq.~(\ref{eqn:MS2}) with the
control term $\varepsilon^2 f_2$ (left panel) and with $\varepsilon^2 f_2+\varepsilon^3 f_3$ (right panel)
given by Eq.~(\ref{eqn:fs}) with $\varepsilon=0.004$, $\psi_0\approx 0.92$ and $\Omega=1$, and with initial
conditions below the surface given by Eq.~(\ref{eqn:surf}).}
\label{fig:6}
\end{figure}

For $\varepsilon\geq 0.003$, the truncated control term $f_2$ is not sufficient to create the barrier.
Therefore, one has to take into account more terms in the series expansion of the control term.
A Poincar\'e section of Hamiltonian~(\ref{eqn:HmsC}) with $H_1$ given by Eq.~(\ref{eqn:MS2}), with the control term $\varepsilon^2 f_2$
and with the control term $\varepsilon^2 f_2+\varepsilon^3 f_3 $ given by Eq.~(\ref{eqn:fs})
for $\varepsilon=0.004$ is represented on Fig.~\ref{fig:6}.

\begin{figure}
\unitlength 1cm
\begin{picture}(7.5,6.3)(0,0)
\put(0,0){\epsfig{file=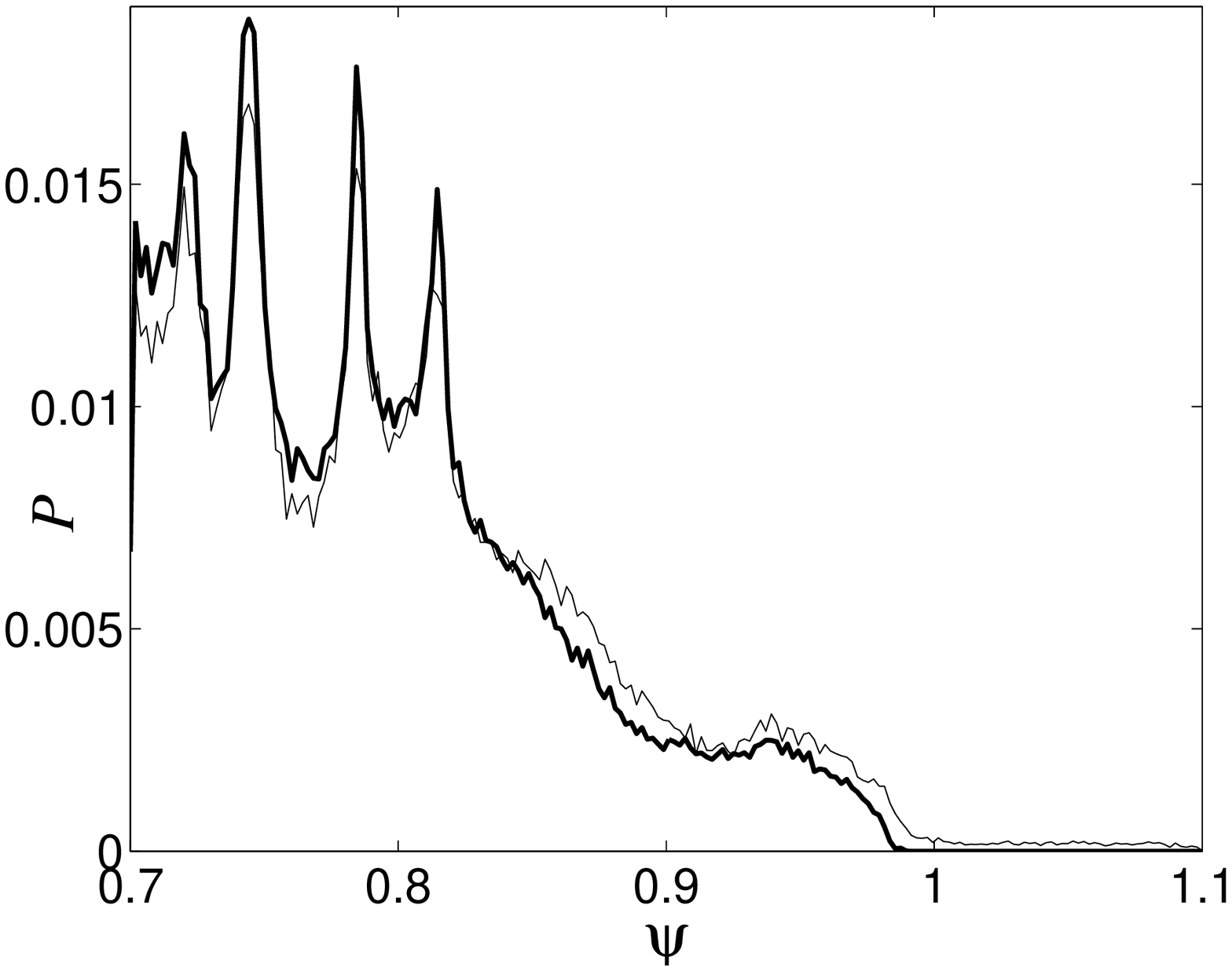,width=7.5cm,height=6.3cm}}
\put(3,3.5){\epsfig{file=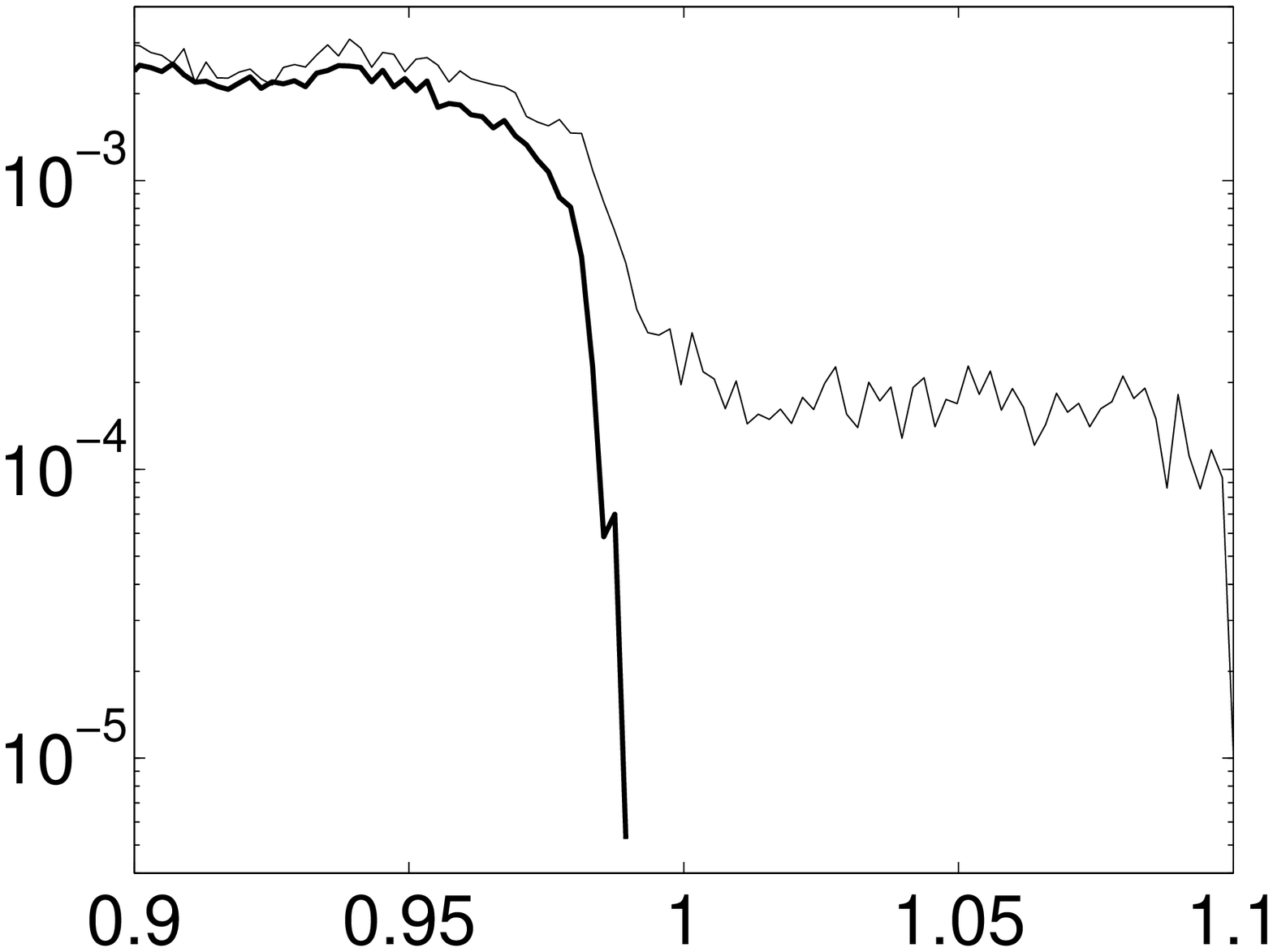,width=4cm,height=2.7cm}}
\end{picture}
\caption{Probability distribution functions of trajectories in the presence of the partial control $f_2$ (dotted line) and the partial control $f_2+f_3$ (bold line).}
\label{fig:7}
\end{figure}

We clearly see that $f_2$ is not sufficient to create an absolute barrier to diffusion in contrast with
the control term obtained with the addition of $f_3$. For higher values of $\varepsilon$,
 more terms of the series can be included if necessary. However we notice that there is still an effective barrier to diffusion which prevents most of the trajectories to diffuse toward the edge. In order to measure the efficiency of the control in the case where the partial control term $f_2$ is not sufficient to create a barrier, we have plotted in Fig.~\ref{fig:7} the probability distribution function of trajectories launched from below the barrier (\ref{eqn:surf}) in the chaotic sea for a fixed elapse of time. This function is averaged over the angles ${\bm\theta}$ so that the barrier is not strictly localized on $\psi_0=0.92$ due to the angle dependence (set by Eq.~(\ref{eqn:surf})). We notice that $f_2$ is still efficient to reduce the diffusion of trajectories since most of the trajectories are such that $\psi\leq 1$.
 
\begin{figure}
\unitlength 1cm
\begin{picture}(15,6.3)(0,0)
\put(0,0){\epsfig{file=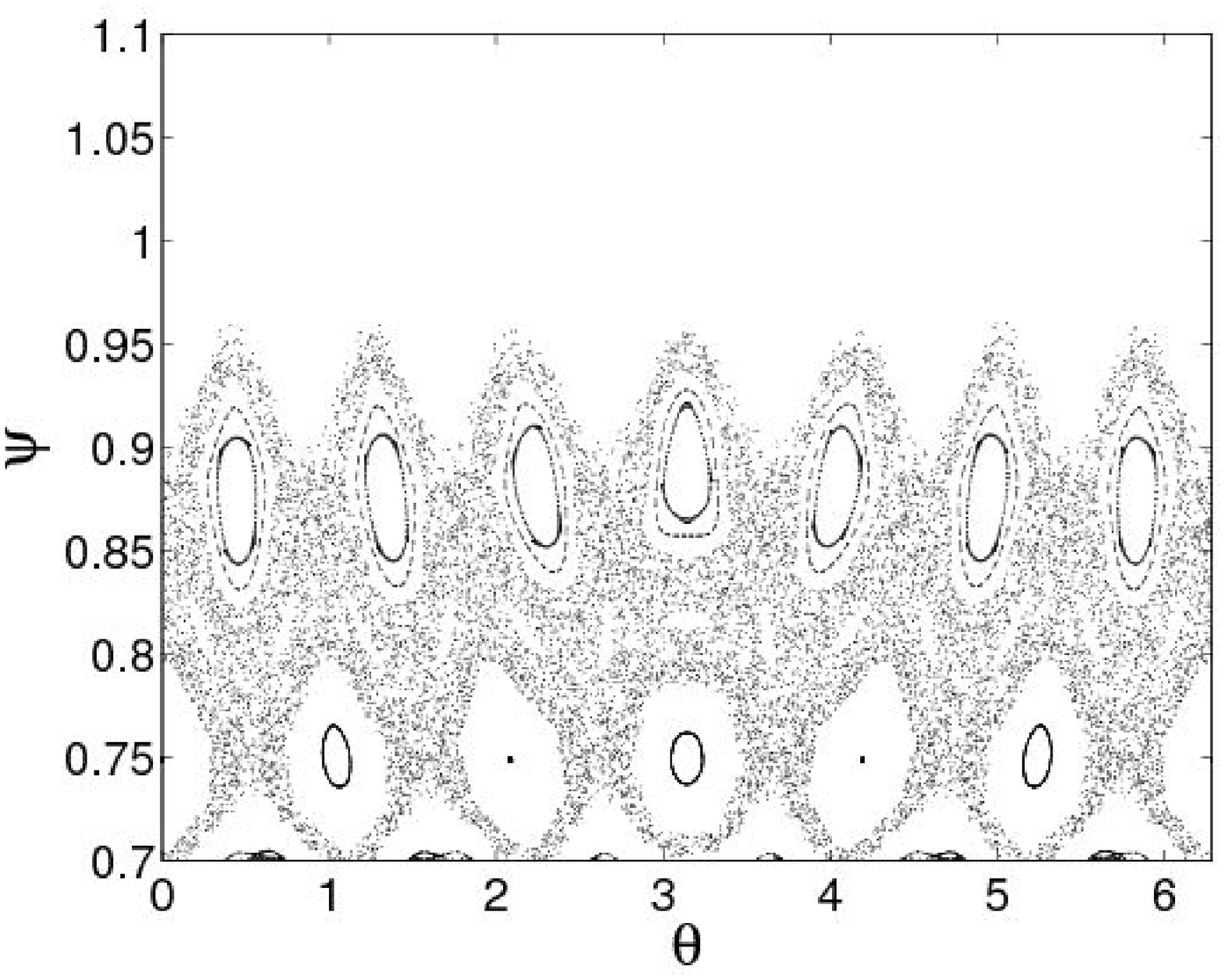,width=7.5cm,height=6.3cm}}
\put(8,0){\epsfig{file=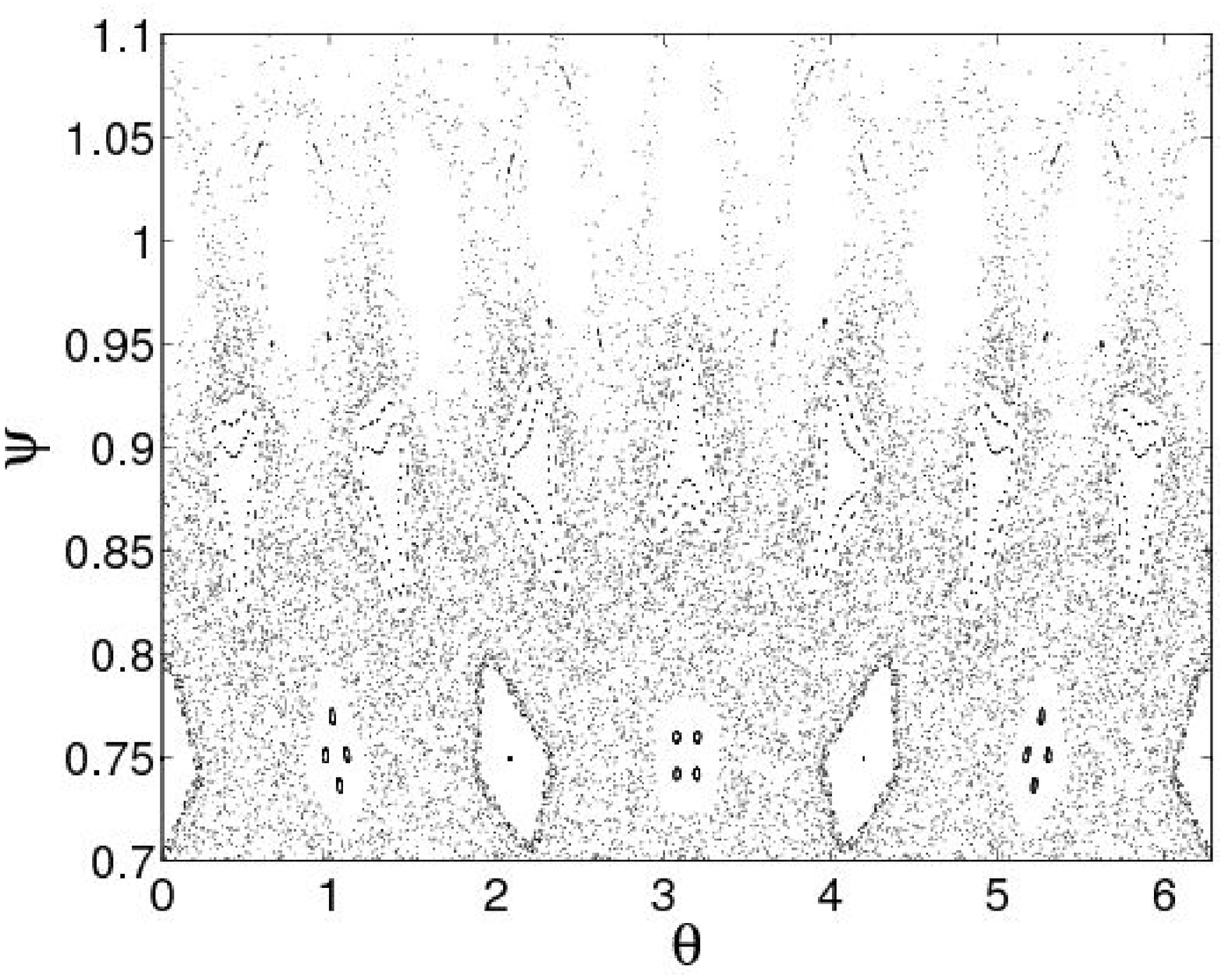,width=7.5cm,height=6.3cm}}
\end{picture}
\caption{Poincar\'e sections with $H_1$ given by Eq.~(\ref{eqn:MS2}) with
the control term given by Eq.~(\ref{eqn:MS2cs}) with $\varepsilon=0.002$ (left panel) and $\varepsilon=0.003$ (right panel) using $\Omega=\Omega_{loc}$ and $\psi_0\approx 0.92$.}
\label{fig:8}
\end{figure}

Moreover, there is the freedom in simplifying the control term by
removing some Fourier components. This can be analyzed when
considering the control term computed for the interaction of two
tearing modes (see Eq.~(\ref{eqn:f2tearing}) in Sec.~\ref{sec:41}), and
generalized to any set of modes, $(m_1,n_1)$ and
$(m_2,n_2)$ with amplitude $A_1$ and $A_2$ characterized by a weak
$\psi$ dependence. In this case, the control term is roughly~: 
\begin{eqnarray*}
\label{eqn:f2tearing_general} f_2(\theta,\varphi)&\approx  &
\Big(-\frac{d(1/q(\psi))}{d\psi}\Big)\Big|_{\psi=\psi_0} \\
&& \times \left(\frac{m_1
A_1\cos(m_1\theta-n_1\varphi)}{m_1\omega-n_1}+\frac{m_2 A_2
\cos(m_2\theta-n_2\varphi)}{m_2\omega-n_2} \right)^2.
\end{eqnarray*}
Such a control term is characterized by four Fourier components,
two are corrections to the amplitude of the original components
and two are coupling terms with angle dependence
$(m_1+m_2)\theta-(n_1+n_2)\varphi$ and
$(m_1-m_2)\theta-(n_1-n_2)\varphi$. Only the first mode is
resonant between the resonant surfaces of main components
and has therefore the largest weight in the control procedure
restoring an invariant torus in its vicinity. This analysis can be
transposed to the control of the stochastic boundary addressed
here, hence such that $n_1=n_2=n$. For instance if one wants to
create a magnetic surface between the island with period 7 and the
one with period 8, the Fourier mode with wave vector $(15,4)$ is
dominant. In the case of a localization between resonant islands
with frequency vector $(m_1,n)$ and $(m_2,n)$, the approximate
control term then reduces to
\begin{equation}
\label{eqn:MS2cs}
f_{2,1}(\theta,\varphi)=f_{m_1 m_2}\cos [(m_1+m_2)\theta-2n\varphi],
\end{equation}
with
\begin{eqnarray*}
f_{m_1 m_2}&=&(-1)^{m_1+m_2}\frac{\sin[(m_1-m_0)\theta_d]}{\pi(m_1-m_0)}\frac{\sin[(m_2-m_0)\theta_d]}{\pi(m_2-m_0)}\psi_0^{(m_1+m_2)/2}\\
&& \times \frac{m_1m_2}{(m_1\omega-n)(m_2\omega-n)}\left(\frac{(m_1+m_2)\omega-2n}{4\psi_0}-\frac{Q'(\psi_0)}{2} \right).
\end{eqnarray*}
In this expression, the $\psi$ dependence of the perturbation
$H_1$ is taken into account leading to a correction in the
magnitude of the control term. Of course, the theorem does no
longer ensure the existence of the invariant torus since this control
term is approximate. However this simplified control term requires
less energy (the ratio between the energy necessary for the
control and the one of the magnetic perturbation is 4\% for
$\varepsilon=0.003$). Furthermore, the strong simplification of
the spectrum of the control term should translate into the design of
the dedicated control coil.  The effect of the simplified control
term of Hamiltonian~(\ref{eqn:MS2}) given by
Eq.~(\ref{eqn:MS2cs}) can be seen on the Poincar\'e section of
Fig.~\ref{fig:8}. It clearly shows that an invariant torus
bounding the motion of magnetic field lines has been created by
this simple control term for $\varepsilon=0.002$. However, when
$\varepsilon$ is increased to $\varepsilon=0.003$ there is no
longer an invariant torus and field lines leak out toward the
$\psi\sim1$ values. The density of points in the Poincar\'e
section of Fig.~\ref{fig:8} is indicative of the existence of a
transport barrier that inhibits the transport at the location of
the invariant torus observed when the full control term is applied. 

\begin{figure}
\epsfig{file=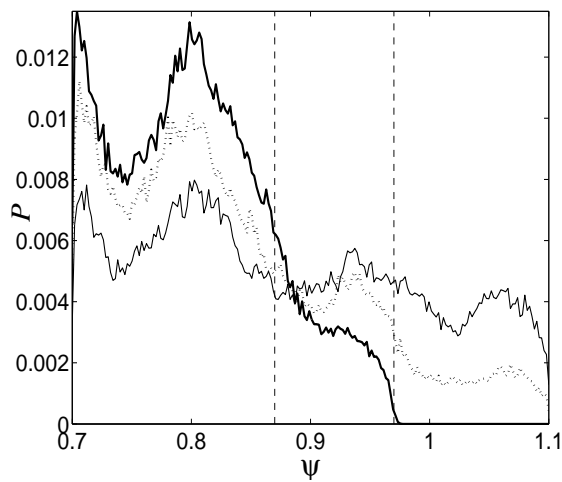,width=7.5cm,height=6.3cm}
\caption{Probability distribution functions of trajectories without control term, with the simplified control term~(\ref{eqn:MS2cs}) (dotted line) and with the control term $f_2$ (bold line). The dashed lines indicate the $\psi$-extension of the created invariant torus. Here $\varepsilon=0.003$.}
\label{fig:9}
\end{figure}

In a similar way as in Fig.~\ref{fig:6}, we have measured the efficiency of the control in the case of Fig.~\ref{fig:8} when $\varepsilon=0.003$ by plotting in Fig.~\ref{fig:9} the probability distribution function (PDF) of trajectories launched from below the barrier (\ref{eqn:surf}) for a fixed time. The bold line represents the PDF with the control term $f_2$ which is sufficient to create a barrier to diffusion when $\varepsilon=0.003$. The thin line is the PDF without control and the dotted line represents the PDF with the simple control given by Eq.~(\ref{eqn:MS2cs}). We notice that the simplified control term is still efficient to reduce the diffusion of trajectories (by a factor 2 for the value of the PDF). In this case, a strongly simplified control term is found to provide the required reduction of transport on the prescribed surface.

Regarding
transport issues in a fusion device this remnant loss of magnetic
confinement can be comparable to the existence of an invariant
torus provided the transport across the barrier along the field
lines is smaller than the turbulent plasma transport
\cite{Garbet2004}.

In this Section, we have shown that a transport barrier consisting
of an invariant torus or a region of reduced field line diffusion
can be generated within a stochastic boundary. This possibility
can prove to be important not only for ergodic divertor
experiments as addressed here but also for other fusion device
such as stellarators, where such stochastic boundaries are
intrinsic to the magnetic equilibrium for a set of configurations
~\cite{stella1,stella2,stella3} as well as Reverse Field Pinches where
restoring the magnetic surfaces in the outermost region is a key
to enhanced performance~\cite{RFP}. Finally, the present results
can also be used to analyze the ELM control scheme addressed on
DIII-D~\cite{DIII-D_2004}. In the latter experiment a very weak
magnetic perturbation is shown to control the large ELMs
associated with the H-mode operation. Analysis of the magnetic
structure have indicated that the magnitude of the perturbation is
comparable to the error field due to toroidal coil misalignment.
There is therefore a possibility that a control of stochastic
transport of the type analyzed here is at work in these
experiments. To investigate such a process one would need to reach
a precise modeling of the ELMs, as proposed in
Ref.~\cite{Marina}, to investigate the impact of such a control
on the ELMs and H-mode pedestal physics. This is clearly beyond
the scope of this paper.

\section{Conclusion and discussion}

We have developed a strategy of localized control of Hamiltonian
systems. This control scheme is original insofar that rather
than compensating for the perturbation driving the stochastic
transport it is based on adding an extra perturbation of lower
magnitude with respect to the initial perturbation in order to
restore an invariant torus. A simple demonstration of this
control scheme is proposed in Sec.~\ref{sec:2}. The application to
magnetic field lines, in Sec.~\ref{sec:3}, allows one to show that it
is possible to create isolated magnetic surfaces within a region
of stochastic field lines.  This prevents magnetic field lines
from diffusing throughout the stochastic region and thus generates
two independent regions separated by a transport barrier. It is
important to stress that in such cases the perturbation leading to
stochastic transport is of order $\varepsilon$ while the control
term is of order $\varepsilon^2$. We note that the control we
propose can be applied in all generality since the
construction is independent of the specific form of the
Hamiltonian. Moreover, this method is also applicable if the
spectrum of the magnetic perturbation is given numerically.
Control for two cases relevant to plasma confinement in magnetic
fusion devices are addressed in Sec.~\ref{sec:4}. 
A rather generic case of the control of stochastic transport generated by two neighboring magnetic islands is provided by the overlap of low mode number tearing modes. In this case, the control term is readily computed and provides a control scheme both in the case where the control term is localized on the area of the restored magnetic surface and in the more relevant case of a control term acting on the whole plasma. The latter control scheme is more in line with a control perturbation generated by external coils.
 Such a control
can prove to be important to confine the plasma current during
the current quench phase that is readily associated with stochastic
transport governed by coupled tearing modes. The second example is
the control of stochastic boundary layers. In such a situation, it
can prove to be important to control the region in contact with
the wall via the stochastic boundary. The calculation of such a
control scheme is performed in the ergodic divertor framework and
can readily be extended to other configurations such as
stellarators. In this case, it is shown that a Fourier truncation
of the control term still provides a control of transport at the
chosen location in the boundary layer. This underlines the
robustness of such a control scheme. It is interesting to note that the control term corresponds to the second harmonics of the spectrum used to generate magnetic perturbations. In a standard analysis, these harmonics are not taken into account. There is therefore a possibility that those parts of the spectrum that are usually neglected, because of their low amplitude and higher harmonics play a role in reducing locally the transport across the stochastic boundary. Going one step further, it is possible, using the ideas presented in this paper, that a weak magnetic perturbation generated by external coils acts as a control term of a boundary magnetic perturbation due to coil misalignment. This issue was raised after the first tests of ELM control on DIII-D with the so-called I coils~\cite{stella1}. The present results indicate that such a counterintuitive property cannot be ruled out without considering the possibility of checking that the controlled perturbation and intrinsic perturbation are not related according to the relationship presented in Eq.~(\ref{eqn:f2}).

\ack
We acknowledge useful discussions with M Pettini and A Wingen.
This work is supported by Euratom/CEA
(contract EUR 344-88-1 FUA F).

\appendix

\section{Existence and regularity of the control term}

In this Appendix, we prove that the control term and the canonical transformation are bounded. It reduces to prove that $\Gamma \partial_{\bm\theta} v$ is bounded. This follows from usual KAM proofs (see for instance Ref.~\cite{kam}) which also shows the existence of invariant tori. The Fourier expansion of $\Gamma \partial_{\bm\theta} v$ writes~:
$$
\Gamma \partial_{\bm\theta} v= \sum_{{\bf k}\in{\mathbb Z}^L\setminus \{{\bf 0}\}}
\frac{{\bf k} v_{\bf k}}
{{\bm \omega}\cdot {\bf k}} {\mathrm e}^{i{\bf k}\cdot {\bm\theta}}.
$$
We assume that ${\bm\omega}$ satisfies a Diophantine condition, i.e.\ there exist $\sigma >0$ and $\tau > L-1$ such that
$$
{\vert {\bm \omega}\cdot {\bf k}\vert}^{-1}\leq \sigma \Vert {\bf k}\Vert^\tau, \qquad \forall {\bf k}\in {\mathbb Z}^L\setminus \{ {\bf 0}\},
$$
where $\Vert {\bf k} \Vert=\sum_i \vert k_i\vert$. 
Moreover, we assume that $v$ is of class $C^{r+\tau+1}$ where $r>1$, i.e.\ bounded for the norm of scalar functions $v({\bm\theta})=\sum_{{\bf k}\in{\mathbb Z}^L}v_{\bf k} {\mathrm e}^{i{\bf k}\cdot {\bm\theta}}$~:
$$
\Vert v\Vert_{r+\tau+1} =\vert v_{\bf 0}\vert+ \sum_{{\bf k}\in{\mathbb Z}^L\setminus \{ {\bf 0}\} } \vert v_{\bf k}\vert \Vert {\bf k}\Vert^{r+\tau+1},
$$
In the same way we define the norm of vectorial functions of class $C^{r}$ as $\Vert {\bf g}\Vert_r=\sum_{i=1}^L \Vert g_i\Vert_r$. Thus we have
$$
\Vert \Gamma \partial_{\bm\theta} v \Vert_r = \sum_{{\bf k}\in{\mathbb Z}^L\setminus \{ {\bf 0}\}} \frac{\Vert {\bf k}\Vert  \vert v_{\bf k}\vert}
{\vert {\bm \omega}\cdot {\bf k}\vert} \Vert {\bf k}\Vert^r.
$$
The Diophantine condition of ${\bm\omega}$ gives
$$
\Vert \Gamma \partial_{\bm\theta} v \Vert_r \leq \sigma \sum_{{\bf k}\in{\mathbb Z}^L\setminus \{ {\bf 0}\}} \Vert {\bf k}\Vert^{\tau+1+r}  \vert v_{\bf k}\vert \leq\sigma \Vert v\Vert_{r+\tau+1}.
$$
This norm is bounded and hence $\Gamma \partial_{\bm\theta} v$ is of class $C^r$. Hence the loss of regularity between $v$ and $\Gamma \partial_{\bm\theta} v$ is the constant $\tau+1$. We notice that we could have also weakened the hypothesis on $(v,\omega)$ into $\Vert \Gamma \partial_{\bm\theta} v\Vert_r <\infty$.

In order to have an estimate on the control term, we denote 
\begin{eqnarray*}
&& \eta_1=\sup_{{\bm\theta}\in{\mathbb T}^L}\Vert \Gamma \partial_{\bm\theta} v \Vert \leq \sigma \Vert v\Vert_{r+\tau+1},\\
&& \eta_2=\frac{1}{\varepsilon}\sup_{\Vert A\Vert \leq \varepsilon \eta_1}\sup_{{\bm\theta}\in{\mathbb T}^L} \max_{i=1,\ldots,L}\vert\partial_{A_i} w({\bf A},{\bm \theta})\vert.
\end{eqnarray*}
If $w$ is given by $w({\bf A},{\bm\theta})=\varepsilon {\bf w}_1({\bm\theta})\cdot {\bf A}+ w_2({\bf A},{\bm\theta})$, we see that
$$
\eta_2\leq \Vert {\bf w}_1\Vert_r+\eta_1 \sup_{\Vert A\Vert \leq \varepsilon \eta_1}\sup_{{\bm\theta}\in{\mathbb T}^L}\max_{i,j} \vert \partial_{A_i} \partial_{A_j} w_2\vert.
$$
The control term given by Eq.~(\ref{eqn:CT}) can be rewritten as
$$
\varepsilon^2 f=\int_0^1 \partial_{\bf A}w(-s\varepsilon \Gamma \partial_{\bm\theta} v,{\bm\theta})\cdot \varepsilon \Gamma \partial_{\bm\theta} v ds.
$$
Thus we have
$$
\sup_{{\bm \theta}\in {\mathbb T}^L}\vert f({\bm\theta})\vert \leq \eta_1\eta_2,
$$
where $\eta_1$ and $\eta_2$ are of order 1, and hence the control term $\varepsilon^2 f$ is of order $\varepsilon^2$. In a very similar way, if we assume that the Hamiltonian is three times differentiable in $({\bf A},{\bm \theta})$, then the derivative of the control term is also bounded and of order $\varepsilon^2$.

\end{document}